\documentclass[preprint,12pt]{elsarticle}

\usepackage{graphicx}

\usepackage{amssymb}

\usepackage{amsmath}
\usepackage{amsfonts}
\usepackage{subfigure}
\usepackage{epstopdf}

\journal{Astroparticle Physics}

\begin{document}

\begin{frontmatter}



\title{Optical Intensity Interferometry with the\\ Cherenkov Telescope Array}


\author[a]{Dainis Dravins}
\author[b]{Stephan LeBohec}
\author[a,c]{Hannes Jensen}
\author[b]{\& Paul D. Nu{\~n}ez}
\author{for the CTA Consortium}

\address[a]{Lund Observatory, Box 43, SE-22100 Lund, Sweden}
\address[b]{Department of Physics and Astronomy, The University of Utah, 115 South 1400 East, Salt Lake City, UT 84112-0830, U.S.A.}
\address[c]{present address: Department of astronomy, Stockholm University,\\ AlbaNova University Center, SE-10691 Stockholm, Sweden}

\begin{abstract}

With its unprecedented light-collecting area for night-sky observations, the Cherenkov Telescope Array (CTA) holds great potential for also optical stellar astronomy, in particular as a multi-element intensity interferometer for realizing imaging with sub-milliarcsecond angular resolution.  Such an order-of-magnitude increase of the spatial resolution achieved in optical astronomy will reveal the surfaces of rotationally flattened stars with structures in their circumstellar disks and winds, or the gas flows between close binaries.  Image reconstruction is feasible from the second-order coherence of light, measured as the temporal correlations of arrival times between photons recorded in different telescopes.  This technique (once pioneered by Hanbury Brown and Twiss) connects telescopes only with electronic signals and is practically insensitive to atmospheric turbulence and to imperfections in telescope optics.  Detector and telescope requirements are very similar to those for imaging air Cherenkov observatories, the main difference being the signal processing (calculating cross correlations between single camera pixels in pairs of telescopes).  Observations of brighter stars are not limited by sky brightness, permitting efficient CTA use during also bright-Moon periods.  While other concepts have been proposed to realize kilometer-scale optical interferometers of conventional amplitude (phase-) type, both in space and on the ground, their complexity places them much further into the future than CTA, which thus could become the first kilometer-scale optical imager in astronomy.  

\end{abstract}

\begin{keyword}

Cherenkov telescopes \sep intensity interferometry \sep Hanbury Brown--Twiss \sep optical interferometry \sep Stars: individual \sep Photon statistics



\end{keyword}

\end{frontmatter}




\section{Resolution frontiers in astronomy}

Many efforts in optical astronomy aim at improving the spatial resolution in order to obtain ever sharper views of our Universe.  Projects include the construction of extremely large telescopes utilizing adaptive optics, or placing instruments in space.  The highest resolution is currently obtained from amplitude (phase-) interferometers which combine light from telescopes separated by baselines up to a few hundred meters.  Since effects of atmospheric turbulence are less severe at longer wavelengths, such instruments are preferentially operated in the near infrared.  Tantalizing results from such facilities show how stellar disks start to become resolved, beginning to reveal stars as a vast diversity of individual objects, although so far feasible only for a small number of the largest ones which extend for tens of milliarcseconds (mas).  More typical bright stars have diameters of only a few mas, requiring interferometry over many hundreds of meters or some kilometer to enable surface imaging.   Using a simple {$\lambda$}/$r$ criterion for the required optical baseline, a resolution of 1 milliarcsecond at {$\lambda$}~500\,nm requires around 100 meters, while 1\,km enables 100\,${\mu}$as.
 
Since we currently are at the threshold of starting to resolve stars as extended objects, a great step forward will be enabled by improving the resolution by just another order of magnitude.  However, since ordinary amplitude interferometers require precisions in both their optics and in the atmosphere above to within a small fraction of an optical wavelength, atmospheric turbulence constrains their operation to baselines not much longer than some 100\,m, especially at shorter visual wavelengths.  

The scientific promise of very long baseline optical interferometry for imaging stellar surfaces has been realized by several \cite{Labeyrie96,Quirr04}, and concepts to circumvent atmospheric turbulence include proposals for large amplitude interferometer arrays in space: {\it{Stellar Imager}} \cite{Carpenter07} and the {\it{Luciola hypertelescope}} \cite{Labeyrie09}, or possibly placed at extreme terrestrial locations such as Dome C in Antarctica \cite{Vakili05}.  However, the complexity and likely cost of these projects make the timescales for their realization somewhat uncertain, prompting searches for alternative approaches.  Although not a complete replacement for the many capabilities of large space-based interferometers, comparable science can begin to be realized very much sooner, and with much less effort, by ground-based intensity interferometry, utilizing large arrays of air Cherenkov telescopes.

\subsection{Intensity interferometry}

Intensity interferometry was pioneered by Robert Hanbury Brown and Richard Q.\ Twiss \cite{HB74}, for the original purpose of measuring stellar sizes, and a dedicated instrument was built at Narrabri, Australia.  It measures temporal correlations of arrival times between photons recorded in different telescopes to observe the second-order coherence of light (i.e., that of intensity, not of amplitude or phase).  The name {\it{intensity interferometer}} is sort of a misnomer: actually nothing is interfering in the instrument.  This name was chosen for its analogy to the ordinary amplitude interferometer, which at that time had similar scientific aims in measuring source diameters.  Two separate telescopes simultaneously measure the random and very rapid intrinsic fluctuations in the light from some particular star.  When the telescopes are placed sufficiently close to one another, the fluctuations measured in the two telescopes are correlated, but when moving them apart, the fluctuations gradually become independent and decorrelated.  How rapidly this occurs for increasing telescope separations gives a measure of the spatial coherence of starlight, and thus the spatial properties of the star.  The signal is a measure of the second-order spatial coherence, the square of that visibility which would be observed in any classical amplitude interferometer.  Spatial baselines for obtaining any given resolution are thus the same as would be required in ordinary interferometry.

The great observational advantage of intensity interferometry (compared to amplitude interferometry) is that it is practically insensitive to either atmospheric turbulence or to telescope optical imperfections, enabling very long baselines as well as observing at short optical wavelengths, even through large airmasses far away from zenith.  Telescopes are connected only with electronic signals (rather than optically), from which it follows that the noise budget relates to the relatively long electronic timescales (nanoseconds, and light-travel distances of centimeters or meters) rather than those of the light wave itself (femtoseconds and nanometers).  A realistic time resolution of perhaps 10 nanoseconds corresponds to 3\,m light-travel distance, and the control of atmospheric path-lengths and telescope imperfections then needs only to correspond to some reasonable fraction of those 3 meters. 

The measured second-order coherence provides the {\it{square}} of the ordinary visibility and always remains positive (save for measurement noise), only diminishing in magnitude when smeared over time intervals longer than the optical coherence time of starlight (due to finite time resolution in the electronics or imprecise telescope placements along the wavefront).  However, for realistic time resolutions (much longer than an optical coherence time of perhaps $\sim$10$^{-14}$\,s), the magnitude of any measured signal is tiny, requiring very precise photon statistics for its reliable determination.  Large photon fluxes (and thus large telescopes) are therefore required; already the flux collectors used in the original intensity interferometer at Narrabri were larger than any other optical telescope at that time. 

Details of the original intensity interferometer at Narrabri and its observing program (mainly measuring angular sizes of hot stars) were documented by Hanbury Brown et al.\ \cite{HB67a,HB67b}, including retrospective overviews \cite{HB74,HB85,HB91}.  The principles are also explained in various textbooks \cite{Lebeyrie06,Saha11,Shih11}.

The original intensity interferometer at Narrabri had two reflecting telescopes of 6.5\,m diameter, formed by mosaics of numerous hexagonal mirrors, providing star images of 12\,arcmin diameter.  Following the completion of that program, the design for a second-generation intensity interferometer was worked out \cite{Davis75,HB79,HB91}.  This larger facility was envisioned to have 12-m diameter telescopes, movable over 2\,km, however it was never realized.  The same physical principles of measuring intensity correlations have since been actively utilized in high-energy particle physics (where also other bosons, i.e., particles with an integer number of their quantum spin, have a tendency to bunch together in a similar way as photons, while electrons and other fermions show the opposite behavior).  In astronomy, however, intensity interferometry has not undergone further development, largely due to its demanding requirements for large optical flux collectors, spread over long baselines, and equipped with fast detectors and high-speed electronics.

\subsection{Air Cherenkov telescopes} \label{1.3}

The parameters of air Cherenkov telescopes are remarkably similar to the requirements for intensity interferometry.  In the Narrabri interferometer, movable telescopes were used to maintain a fixed baseline while tracking a source across the sky.  Nowadays, electronic time delays can compensate for the different arrival times of a wavefront to different telescopes in fixed positions. 

The most remarkable potential comes from the Cherenkov Telescope Array \cite{Actis11,CTA12} which foresees a total of 50--100 telescopes with differently sized apertures between about 5 and 25 meters, distributed over an area of 2--3\,km{$^2$}.  Such a large array permits an enormous number of optical baseline pairs to be synthesized, enabling measurements of angular scales between milli- and microarcseconds.  The potential of using such arrays for intensity interferometry has indeed been noticed by several authors \cite{deWit08,LeBohec08a,LeBohec06}.  Within the CTA project, a task group was set up to specify how to enable it for also such uses.  If a baseline of 2\,km could be utilized at {$\lambda$}~=\,350\,nm, resolutions would approach 30\,{$\mu$}as, an unprecedented spatial resolution in optical astronomy.  Such numbers are challenged only by radio interferometers operating between Earth and antennas in deep space \cite{Kardshev09}, or possibly by futuristic X-ray interferometers \cite{MAXIM11}.

\section{Principles of intensity interferometry}
\label{iiprinciple}

\begin{figure}
\centering
\includegraphics[width=9cm]{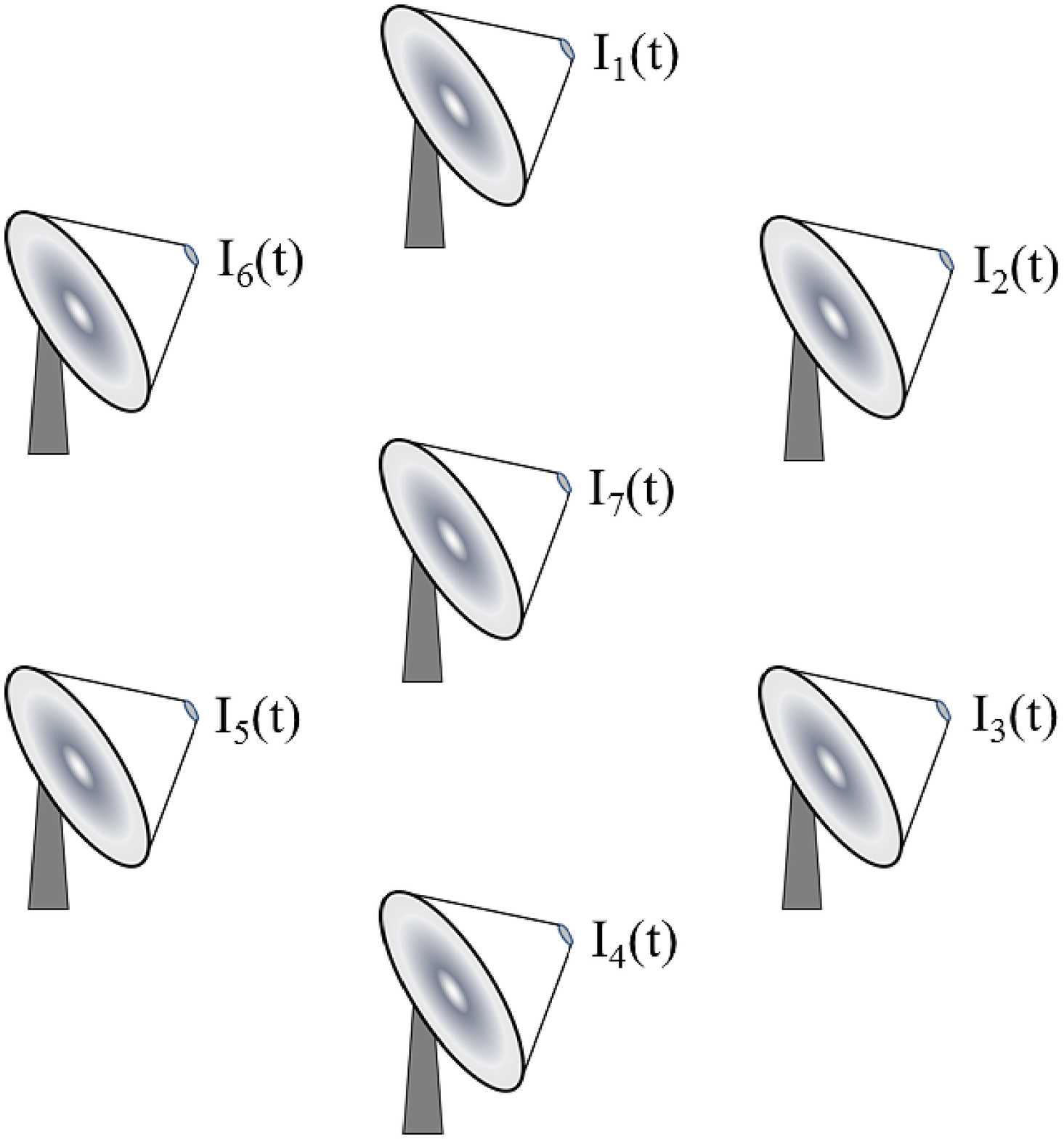}
\caption{ Principle of a multi-element stellar intensity interferometer.  Several telescopes observe the same source, simultaneously recording its rapidly fluctuating optical light intensity $I_n(t)$.  Cross correlations of the intensity fluctuations are measured between different pairs of telescopes: $\langle I_1(t) I_2 (t) \rangle$, $\langle I_1(t) I_3 (t) \rangle$, $\langle I_1(t) I_4 (t) \rangle$, $\langle I_2(t) I_3 (t) \rangle$, $\langle I_2(t) I_4 (t) \rangle$, etc.  These yield a measure of the second-order spatial coherence of light, from which an image of the source can be deduced, with an angular resolution corresponding to the optical diffraction over the projected baseline distance between each pair of telescopes.  Numerous telescopes enable a very large number of baselines to be synthesized, permitting a high-fidelity reconstruction of the source image.  Telescopes distributed over also km-long baselines enable an angular resolution so far unprecedented in optical astronomy. }
\label{principles}
\end{figure}

In its simplest form, an intensity interferometer consists of two optical telescopes or light collectors, each with a photon detector feeding one channel of a signal processor for temporally cross correlating the light-intensity signals from the two telescopes.  The intensities measured at detectors 1 and 2 are the respective values of the electric light-wave amplitude times its complex conjugate, averaged over some time interval corresponding to the signal bandwidth of the detectors and associated electronics:

\begin{equation}
  \langle I(t) \rangle = \langle E(t)E^*(t) \rangle
\end{equation}


where $^*$ marks complex conjugate and $\langle~\rangle$ denotes averaging over time.  The intensities  measured in the two telescopes are cross correlated:

\begin{equation}
  \langle I_1(t) I_2(t) \rangle = \langle E_1(t)E^*_1(t) \cdot E_2(t)E^*_2(t) \rangle
\end{equation}

This expression can be expanded by dividing the complex field amplitudes into their real and imaginary parts.  Here one must make an assumption that is fundamental to the operation of an intensity interferometer: the light must be chaotic, i.e., with a Gaussian amplitude distribution; also called thermal- or maximum-entropy light \cite{BachorRalph04,Foellmi09,Loudon00,Shih11}.  To a good approximation this applies to all `ordinary' light sources (but not necessarily to nonthermal ones such as lasers).  Such light may well be quasi-monochromatic, as long as the light waves undergo random phase shifts, so that intensity fluctuations result on timescales corresponding to the optical coherence time.  For chaotic light, the real and imaginary parts of $E_1$ and $E_2$ are Gaussian random variates, i.e., the values of $E_1$ and $E_2$ measured at different times can be treated as random variables obeying a normal distribution.  Then the Gaussian moment theorem applies, which relates all higher-order correlations of Gaussian variates to products of their lower-order correlations (described in detail by Mandel \& Wolf \cite{Mandel95}).   It is then possible to show \cite{Lebeyrie06} that, for linearly polarized light:

 \begin{equation}
  \langle I_1(t) I_2 (t) \rangle = \langle I_1 \rangle \langle I_2 \rangle (1 + |\gamma_{12}|^2)
\end{equation}

where $\gamma_{12}$ is the mutual coherence function of light between locations 1 and 2, the quantity measured in ordinary amplitude interferometers. 

Defining the intensity fluctuations $\Delta I$ as:

\[
\Delta I_1(t) = I_1(t) - \langle I_1 \rangle
\qquad \Delta I_2(t) = I_2(t) - \langle I_2 \rangle ,
\]

one obtains:

\begin{equation}
\label{intcorr4}
\langle \Delta I_1(t) \Delta I_2(t) \rangle = \langle I_1 \rangle \langle I_2 \rangle |\gamma_{12}|^2,
\end{equation}

since $\langle \Delta I \rangle = 0$.

An intensity interferometer thus measures $|\gamma_{12}|^2$ with a certain electronic time resolution.  This quantity remains positive irrespective of atmospheric or optical disturbances although -- since realistic time resolutions do not reach down to optical coherence times -- it may get strongly diluted relative to the full value it would have had in the case of a hypothetical `perfect' temporal resolution (shorter than the light-wave period).  For realistic values of nanoseconds, this dilution typically amounts to several orders of magnitude and thus the directly measurable excess correlation becomes quite small.  This is the reason why very precise photon statistics are required, implying large flux collectors.

\section{Optical aperture synthesis}
\label{apertsynt}

The original intensity interferometer at Narrabri used two telescopes, movable on railroad tracks, which could be positioned at different separations $r$, to deduce angular sizes of stars from the observed function $|\gamma_{12}(r)|^2$, analogous to what can be measured with a two-element amplitude interferometer.  Systems with multiple telescopes and different baselines (Figure \ref{principles}) enable correspondingly more complete image reconstructions.  Techniques for interferometric imaging and aperture synthesis were first developed for radio telescopes \cite{Taylor99,Thompson01}, but have since been elaborated also for the optical \cite{Glindemann11, Lebeyrie06, Saha11}.  Here we recall the basics:

The separation vector between a pair of telescopes in a plane perpendicular to the line of observation, the $(u,v)-$plane, is $\mathbf{r_1-r_2}$, so that for an optical wavelength $\lambda$, $\mathbf{r_1-r_2} = (u \lambda, v \lambda)$.  If the telescopes are not in such a plane, also a third coordinate enters: the time-delay $w$ for the propagation of light along the line of sight to the source; $\mathbf{r_1-r_2} = (u \lambda, v \lambda, w)$.

With the angular coordinate positions of the target $(l, m)$, one can deduce the following expression for the correlation function $\Gamma_{12} = \langle E(\mathbf{r}_1) E^*(\mathbf{r}_2) \rangle$:

\begin{equation}
\label{vancittert1}
\Gamma(u,v) = \iint I(l, m) e^{-2 \pi i (ul+vm)} dl dm.
\end{equation}

This equation represents the van Cittert-Zernike theorem, equating the quantity measured by an [amplitude] interferometer for a given baseline to a component of the Fourier transform of the surface intensity distribution of the source.  This Fourier transform can be inverted:

\begin{equation}
\label{vancittert}
I_{\nu}(l,m) = \iint V(u, v) e^{2 \pi i (ul+vm)} du dv,
\end{equation}

where $V(u,v)$ equals the normalized value of $\gamma(u,v)$.  Thus, by using multiple separations and orientations of interferometric pairs of telescopes, one can sample the $(u,v)-$plane and reconstruct the source image with a resolution equal to that of a telescope with a diameter of the longest baseline.  This is the technique of aperture synthesis.

In intensity interferometry, however, an additional complication enters in that the correlation function for the electric field, $\gamma_{12}$, is not measured directly, but only the square of its modulus, $|\gamma_{12}|^2$.  Since this does not preserve phase information, the {\it{direct}} inversion of the above equation is not possible.

This limitation will be removed in intensity interferometry carried out with larger telescope arrays.  For CTA, with some 50 or more elements, the possible number of baselines between telescope pairs becomes enormous; $N$ telescopes can form $N(N-1)/2$  baselines, reaching numbers in the thousands (even if possibly periodic telescope locations might make several of them redundant).  Since such telescopes are fixed on the ground, the projected baselines trace out curves in the $(u,v)-$plane, as a source moves across the sky.  With proper signal handling, all successive measures of $|\gamma_{12}|^2$ can be allocated to their specific $(u,v)-$coordinates, producing a highly filled $(u,v)-$plane, with a superior coverage of projected orientations across the source image.  As will be discussed below, such complete data coverage indeed enables reconstruction of the phases of the Fourier components, and thus permits full two-dimensional image reconstructions (although the completeness of such coverage depends on how the source moves across the sky and thus on, e.g., whether it is located near the celestial equator or close to its poles).

For large numbers of telescopes, another advantage of intensity interferometry becomes obvious.  Since telescopes connect only with electronic signals, there is in principle no loss of data when synthesizing any number of baselines between any pairs of telescopes: the digital signal from each telescope is merely copied electronically.   By contrast, amplitude interferometry in the optical (as opposed to radio) requires optical beams of actual starlight between telescopes since the very high optical frequency (combined with rapid phase fluctuations in chaotic light) precludes its amplification with retained phase information.  In order to obtain the many baselines needed for efficient aperture synthesis (such as realized in radio), starlight from each telescope is split and sent to beam combiners to interfere with the light from other telescopes, each combination with its own delay-line system.  While such ambitious arrangements can be made for a moderate number of telescopes \cite{Creech10}, the complexity (and the dilution of light between different baselines) rapidly increases if any greater number of telescopes would be engaged.

\section{The Cherenkov Telescope Array}
\label{cherenkov}

\begin{figure}
\centering
\includegraphics[width=9cm, angle=90]{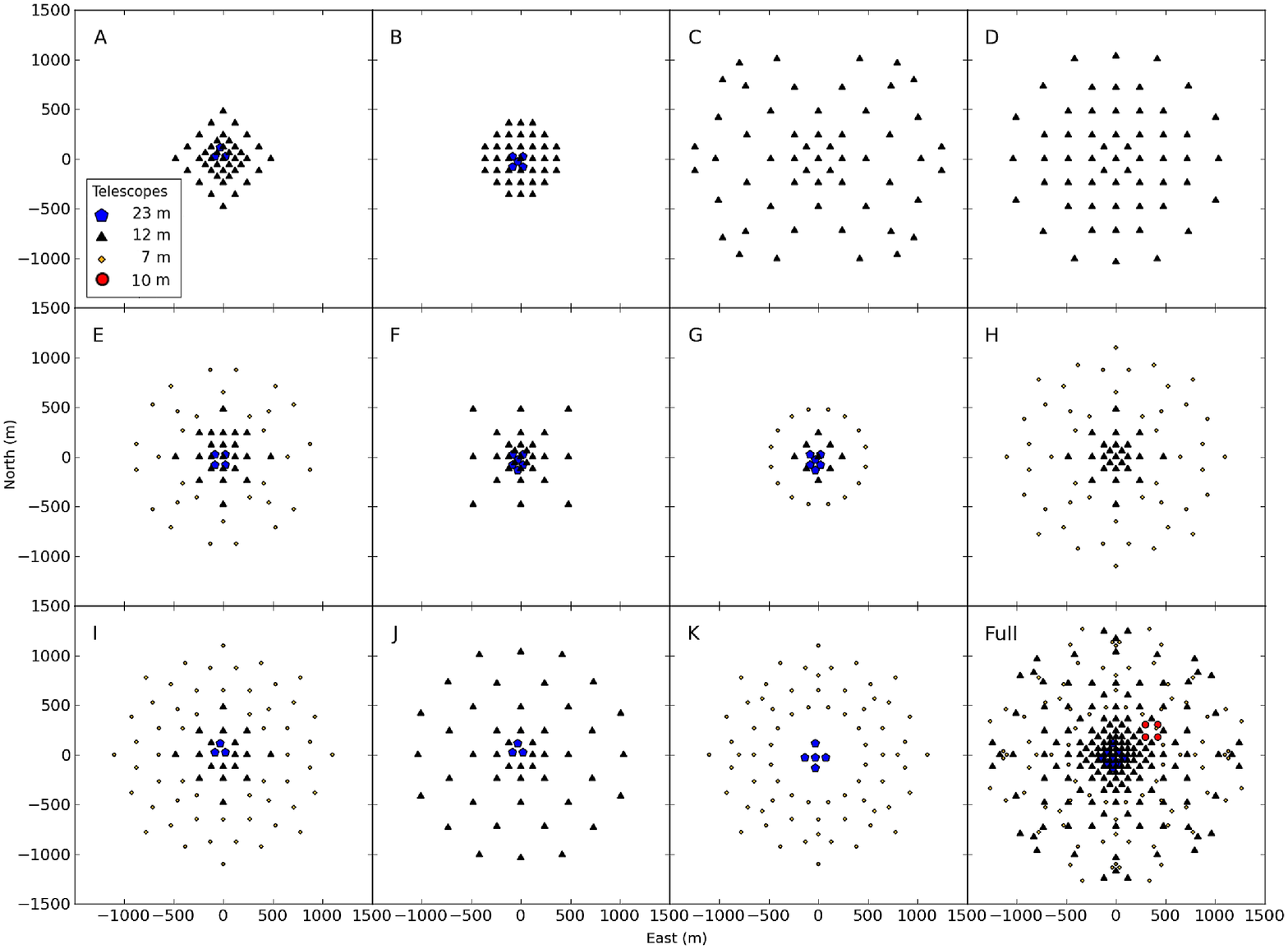}
\caption{Different telescope array layouts evaluated within the CTA design study, and also considered for their suitability to intensity interferometry. Each of the configurations labeled A through K is a subset of the all-encompassing hypothetical large array shown at bottom right.  In this work, configurations B, D and I were selected as representative for three different classes of array geometry.}
\label{configs}
\end{figure}

CTA is envisioned to have on the order of 50-100 telescopes with various apertures between about 5 and 25 meters, with currently favored configurations reaching an edge-to-edge distance of some 2\,km.  A number of candidate array layouts for the CTA were considered within its design study \cite{Actis11, Bernlohr08, CTA12, Hermann10}: Figure \ref{configs},  of which examples representing qualitatively different types of layouts are in Table \ref{configurationstable}.  For interferometry, large telescope separations (long baselines) measure high-frequency Fourier components, corresponding to small structures on the target, while short baselines sample the low frequencies.  For an Earth-bound interferometer (in a plane perpendicular to the line of observation) with a baseline $\mathbf{B} = (B_{\mathrm{North}}, B_{\mathrm{East}})$ the associated coordinates in the Fourier $(u,v)-$plane are $(u,v) = \frac{1}{\lambda}(B_{\mathrm{North}}, B_{\mathrm{East}})$.

\begin{table}[b]
\caption{Properties of the three examined array layouts (B, D, and I in Figure \ref{configs}) from the CTA design study \cite{Actis11}.   $N$ is the number of telescopes, $A$ is the light-collection area of each type of telescope, $b$ is the number of unique baselines available, $B_{min}, B_{max}$ indicate the range of baselines for observations in zenith. The corresponding range of angular diameters in milliarcseconds $(1.22 \lambda/r)$ for observations at $\lambda$~400\,nm is indicated by  $\theta_{min}, \theta_{max}$. }
\begin{center}
\begin{tabular}{lrrrrr|}
Array & N & A [m$^2$] & $b$ & $B_{min}, B_{max}$ [m] & $\theta_{min}, \theta_{max}$ [mas]\\
\hline \hline
B & 42 & 113, 415 & 253 & 32, 759 & 0.13, 3.2 \\
D & 57 & 113 & 487 & 170, 2180 & 0.05, 0.6 \\
I & 77 & 28, 113, 415 & 1606 & 90, 2200 & 0.05,  1.13	
\end{tabular}
\end{center}
\label{configurationstable}
\end{table}

\begin{figure}
\centering
\includegraphics[width=8cm]{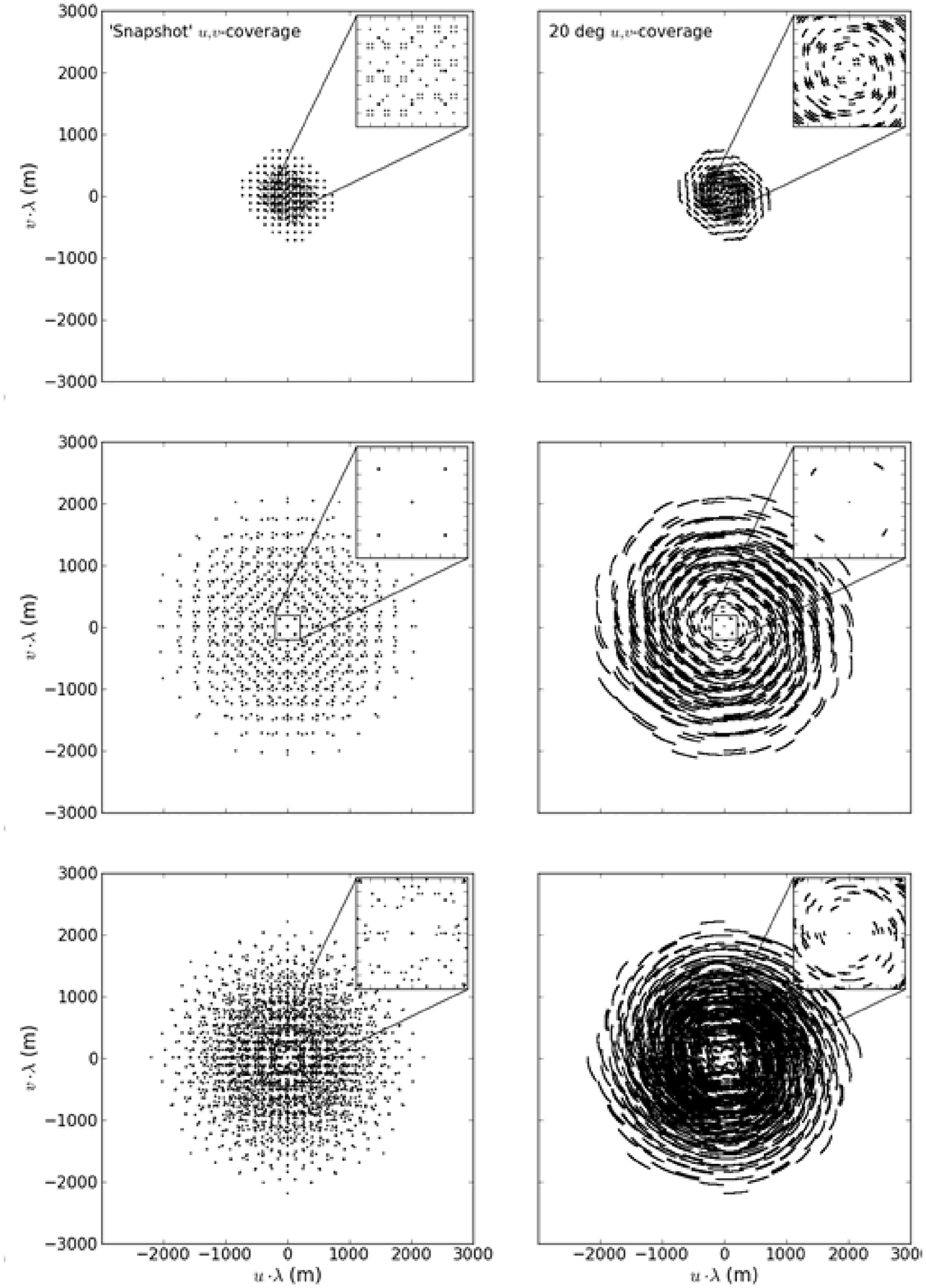}
\caption{Coverage of the interferometric $(u,v)-$plane for three types of evaluated CTA layouts (top to bottom: B, D and I of Figure \ref{configs}).  Left: $(u,v)-$plane 
coverages at one instant in time, for a star observed in the zenith. Upper right-hand squares expand the central 400{$\times$}400\,m area.  Right: $(u,v)-$plane coverages for a star moving from the zenith through 20 degrees to the west.  The numerous telescopes enable a huge number of baseline pairs which largely fill the entire $(u,v)-$plane.}
\label{uv_cover}
\end{figure}

For stationary telescopes, the projected baselines, $\mathbf{B}_p$, will change while the target of observation moves across the sky, with each telescope pair tracing out an ellipse in the Fourier plane according to the following expression \cite{Segran07}:

\begin{scriptsize}
\begin{equation}
\label{rotationsynth}
\begin{pmatrix}
u \\
v \\
w
\end{pmatrix}
=
\frac{1}{\lambda} \mathbf{B}_p = 
\frac{1}{\lambda} \begin{pmatrix}
-\sin l \sin h & \cos h & \cos l \sin h\\
\sin l \cos h \sin \delta + \cos l \cos \delta & \sin h \sin \delta & -\cos l \cos h \sin \delta + \sin l \cos \delta \\
-\sin l \cos h \cos \delta + \cos l \sin \delta & -\sin h \cos \delta & \cos l \cos h \cos \delta + \sin l \sin \delta
\end{pmatrix}
\begin{pmatrix}
B_{north} \\
B_{east}\\
B_{up}
\end{pmatrix}
\end{equation}
\end{scriptsize}

where $l$ is the latitude of the telescope array, and $\delta$ and $h$ are the declination and hour angles of the star. The $w$ component corresponds to the time delay in the wavefront arrival time between the two telescopes (dependent on also the elevation difference of the telescopes, $B_{up}$).  The extensive coverage of the $(u,v)-$plane that results from the Earth's rotation enables the synthesis in software of a very large telescope and -- of course -- is the very principle used in much of radio interferometry.

Figure \ref{uv_cover} illustrates these capabilities for three among the potential layouts considered for CTA, here taken as examples of qualitatively different telescope arrangements.  One is a compact configuration; another a sparse and rather uniform one; and a third has telescopes of different sizes grouped with successively different spacings.  The latter type of layout seems to lie close to those currently favored for CTA in general, and is also the most capable one for interferometry. As seen in Figure \ref{uv_cover}, already short observations of just an hour or so, may cover much of the $(u,v)-$plane (and the coverage can be increased by observing at different wavelengths).

\section{Signal-to-noise ratios in intensity interferometry} \label{5}

For one pair of telescopes, the signal-to-noise ratio is given by \cite{HB74,Twiss69}:

\begin{equation}
\label{signalnoise}
(S/N)_{RMS} = A \cdot \alpha \cdot n \cdot |\gamma_{12}(\mathbf{r})|^2 \cdot \Delta f^{1/2} \cdot (T/2)^{1/2} 
\end{equation}

where $A$ is the geometric mean of the areas (not diameters) of the two telescopes; $\alpha$ is the quantum efficiency of the optics plus detector system; $n$ is the flux of the source in photons per unit optical bandwidth, per unit area, and per unit time; $|\gamma_{12}(\mathbf{r})|^2$ is the second-order coherence of the source for the baseline vector $\mathbf{r}$, with $\gamma_{12}(\mathbf{r})$ being the mutual degree of coherence.  $\Delta f$ is the electronic bandwidth of the detector plus signal-handling system, and $T$ is the integration time.

Most of these parameters depend on the instrumentation,  but $n$ depends on the source itself, being a function of its radiation temperature.  For a given number of photons detected per unit area and unit time, the signal-to-noise ratio is better for sources where those photons are squeezed into a narrower optical band.   This property implies that (for a flat-spectrum source) the S/N is {\it{independent of}} the width of the optical passband, whether measuring only the limited light inside a narrow spectral feature or a much greater broad-band flux.  Although perhaps somewhat counter-intuitive, the explanation is that realistic electronic resolutions of nanoseconds are very much slower than the temporal coherence time of broad-band light (perhaps 10$^{-14}$\,s).  While narrowing the spectral passband does decrease the photon count rate, it also increases the temporal coherence by the same factor, canceling the effects of increased photon noise.  This property was exploited already in the Narrabri interferometer \cite{HB70} to identify the extended emission-line volume from the stellar wind around the Wolf-Rayet star $\gamma^2$~Vel.  The same effect could also be exploited for increasing the signal-to-noise by observing the same source simultaneously in multiple spectral channels, a concept foreseen for the once proposed successor to the original Narrabri interferometer \cite{Davis75,HB79, HB91}.

\section{Simulated observations in intensity interferometry} \label{6}

To obtain quantitative measures of what can be observed using realistic detectors on Cherenkov telescopes, a series of simulations were carried out.

\subsection{Numerical simulations} \label{6.1}

An intensity interferometer using two photon-counting detectors $A$ and $B$ and a digital correlator measures the squared modulus of the complex degree of coherence of the light:

\begin{equation}
\label{intcorr}
|\gamma|^2 = \frac{\langle \Delta I_1 \Delta I_2 \rangle}{\langle I_1 \rangle \langle I_2 \rangle}
\end{equation}

or, in a discrete form:

\begin{equation}
\label{disccorr}
g^{(2)} = \frac{N_{AB}}{N_A N_B} N,
\end{equation}

where $N_A$ and $N_B$ are the number of photons detected in $A$ and $B$ respectively, $N_{AB}$ is the number of joint detections (i.e., the number of time intervals in which both detectors record a photon), and $N$ is the number of sampled time intervals.  Since a strict Monte-Carlo simulation would be computationally very demanding, a simplified procedure was used by generating random numbers $N_A$, $N_B$ and $N_{AB}$, and inserting these into Eq. \eqref{disccorr}. These are Poisson-distributed random variables with mean values $\mu_A = P_A\cdot N$, $\mu_B = P_B\cdot N$ and $\mu_{AB} = P_{AB}\cdot N$. Here, $P_A$ and $P_B$ are the probabilities of detecting a photon in $A$ and $B$ respectively, within a small time interval $\Delta t$, and $P_{AB}$ is the probability of a joint detection within $\Delta t$.

These probabilities can be written out in terms of variables depending only on the instrumentation and the target of study:
\begin{align}
P_A &= \alpha_A \langle I_A \rangle \Delta t \\
P_B &= \alpha_B \langle I_B \rangle \Delta t \\
P_{AB} &= P_A P_B + \alpha_A \alpha_B \langle I_A \rangle \langle I_B \rangle |\gamma_{AB}|^2 {\tau_c}{\Delta t} 
\end{align}

\begin{figure}
\centering
\includegraphics[width=5cm, angle=90]{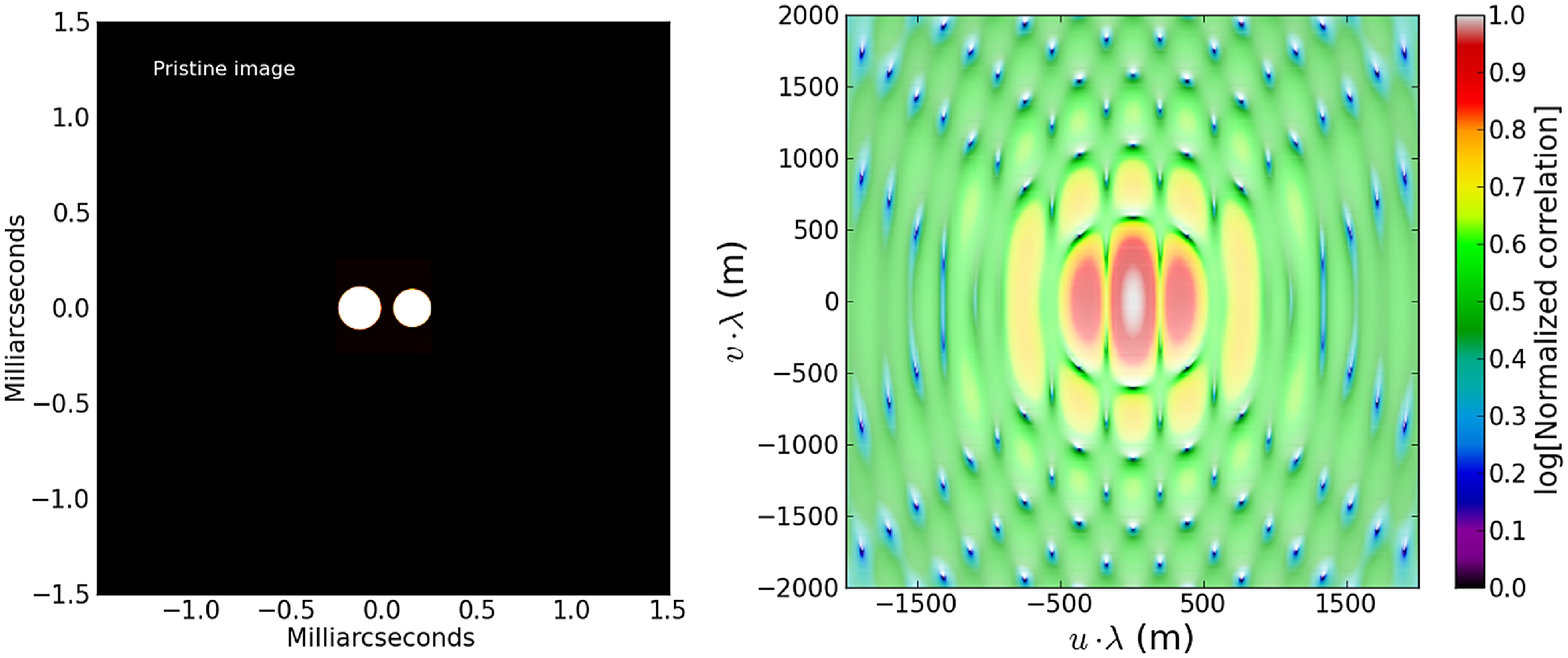}
\caption{Image of a close binary star with 200 and 150\,$\mu$as diameter components, used to simulate observations, and the (logarithmized) magnitude of its Fourier transform.  This noise-free pattern is what would be measured by a perfect interferometer of projected size 2000{$\times$}2000\,m.  Corresponding patterns in later figures cover only some part of this  $(u,v)-$plane (due to finite extent of the telescope array on the ground), and become noisy for fainter sources and finite integration times. }
\label{input}
\end{figure}

Here $\alpha$ denotes the quantum efficiency of the detectors, $\langle I \rangle$ is the mean light intensity, $\tau_c$ is the coherence time of the light (determined by the wavelength and optical passband) and $\gamma_{AB}$ is the degree of optical coherence (proportional to the Fourier transform of the target image, assuming telescope sizes to be small compared to the spatial structure in this transform).   Such simulations were carried out for various telescope-array configurations and for various assumed sources.  Here, examples are shown for a close binary star with components taken as uniform disks of diameters 200 and 150\,$\mu$as.  Both the pristine original image and its pristine Fourier transform in the $(u,v)-$plane are shown in Figure \ref{input}.  Across the Fourier plane, the magnitude of various patterns varies greatly.  To enhance the visibility of also fainter structures (and later to better see the effects of noise), the Fourier-plane figures use a logarithmic scaling and a shading to enhance the contrast  (the exact numerical values of the measured correlations are not significant in this context).

Also results from the simulated observations are mostly given as such Fourier-plane images rather than full image reconstructions.  The simulated observations produced values at many different discrete locations in the $(u,v)-$plane, which were used in a linear interpolation to obtain the Fourier magnitude over a regular grid.  This image format makes the effects of noise and changing telescope arrangements easier to interpret since it is independent of the performance of algorithms for image reconstruction or data analysis.  As discussed below, optimal image reconstruction is a developing research topic of its own.  Even though reconstructed images do reflect the capability of the simulated telescope array, some reconstructions are still limited by the algorithms used.  By contrast, the information recovered in the $(u,v)-$plane is independent of such algorithm performance.

\subsection{Limiting stellar magnitudes}

The question of how faint sources that can be usefully observed has been examined \cite{Dravins12,LeBohec06}, with the conclusion that a conservative practical limit for {\it{two-dimensional}} imaging with a large array of the CTA type is around m$_V$=\,6.  However, if only some {\it{one-dimensional}} measure would be sought (e.g., a stellar diameter or limb darkening), the data can be averaged over all position angles, and the limiting magnitude will become somewhat fainter.  In any case, there are thousands of stars bright enough to be observable.

\section{Imaging with intensity interferometry}
\label{reconstruction}

An intensity interferometer directly measures only the absolute magnitudes of the respective Fourier transform components of the source image that cover the $(u,v)-$plane, while the phases are not directly obtained.  Such Fourier magnitudes can well be used by themselves to fit model parameters such as stellar diameters, stellar limb darkening, binary separations, circumstellar disk thicknesses, etc., but two-dimensional images cannot be directly computed from the van~Cittert-Zernicke theorem, Eq.\eqref{vancittert}.  However, a multi-component interferometer offers numerous baselines, and gives an extensive coverage of the $(u,v)-$plane, and it is already intuitively clear that the information contained there must place rather stringent constraints on the source image.

\subsection{Phase reconstruction}
\label{phase_reconstruction}

A number of techniques have been developed for recovering the phase of a complex function when only its magnitude is known.  Methods specifically intended for intensity interferometry have been worked out for one \cite{Holmes04} or two dimensions \cite{Holmes10}.  Once a sufficient coverage of the Fourier plane is available, and phase recovery has been performed, image reconstruction becomes straightforward.

\begin{figure}
\centering
\includegraphics[width=8cm, angle=90]{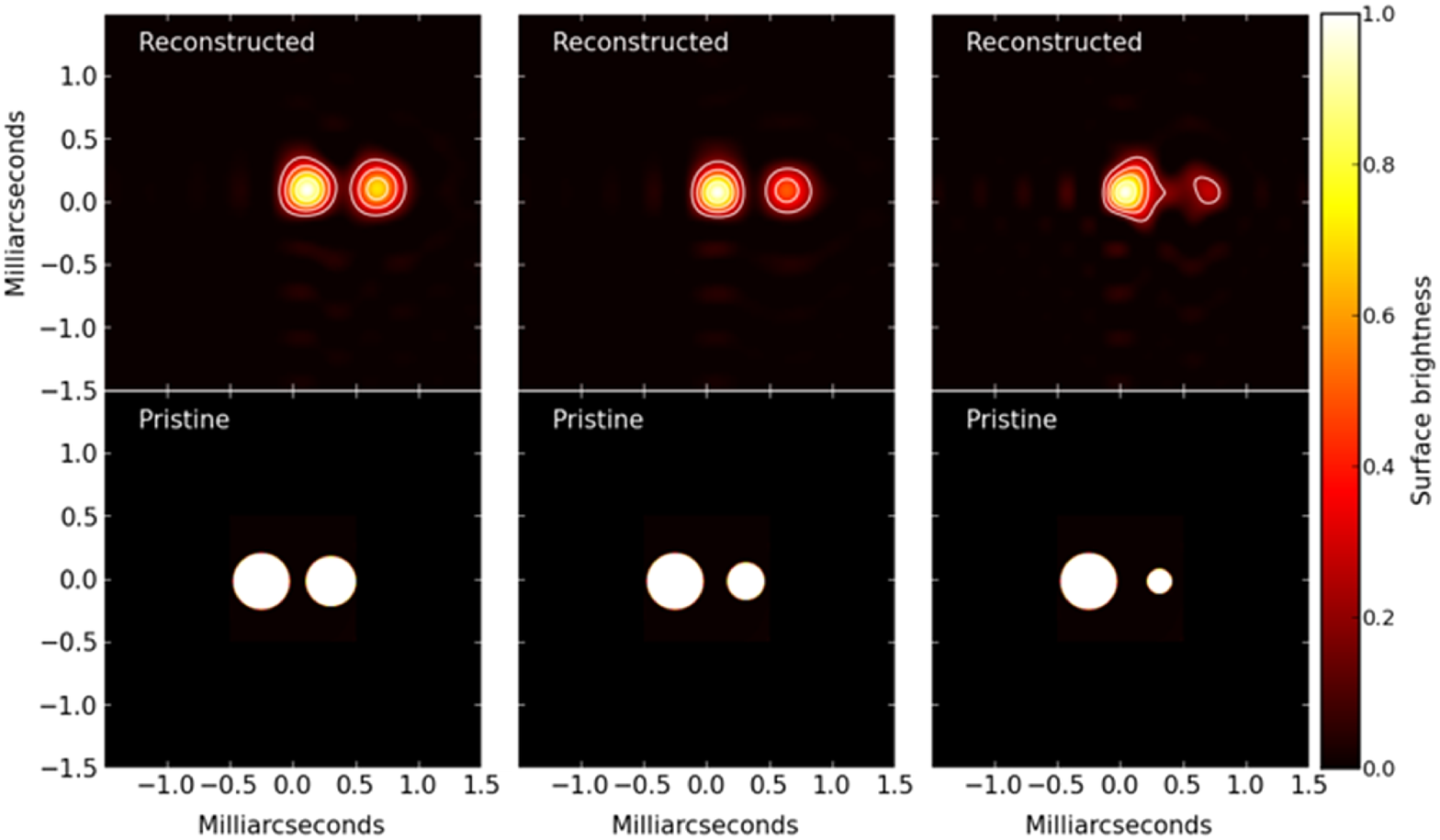}
\caption{Reconstructed images of binary stars (with varying diameter of the secondary) from simulated CTA observations.  These simulations were for the array layout B (Figure \ref{configs}), for sources assumed to have visual magnitude $m_V$=\,3, and effective temperature $T_{eff}=7000$\,K.  The assumed pristine images are shown below while the corresponding $(u,v)-$plane coverage is in Figure \ref{binsepFT}.  }
\label{binsep}
\end{figure}

\begin{figure}
\centering
\includegraphics[width=4.5cm, angle=90]{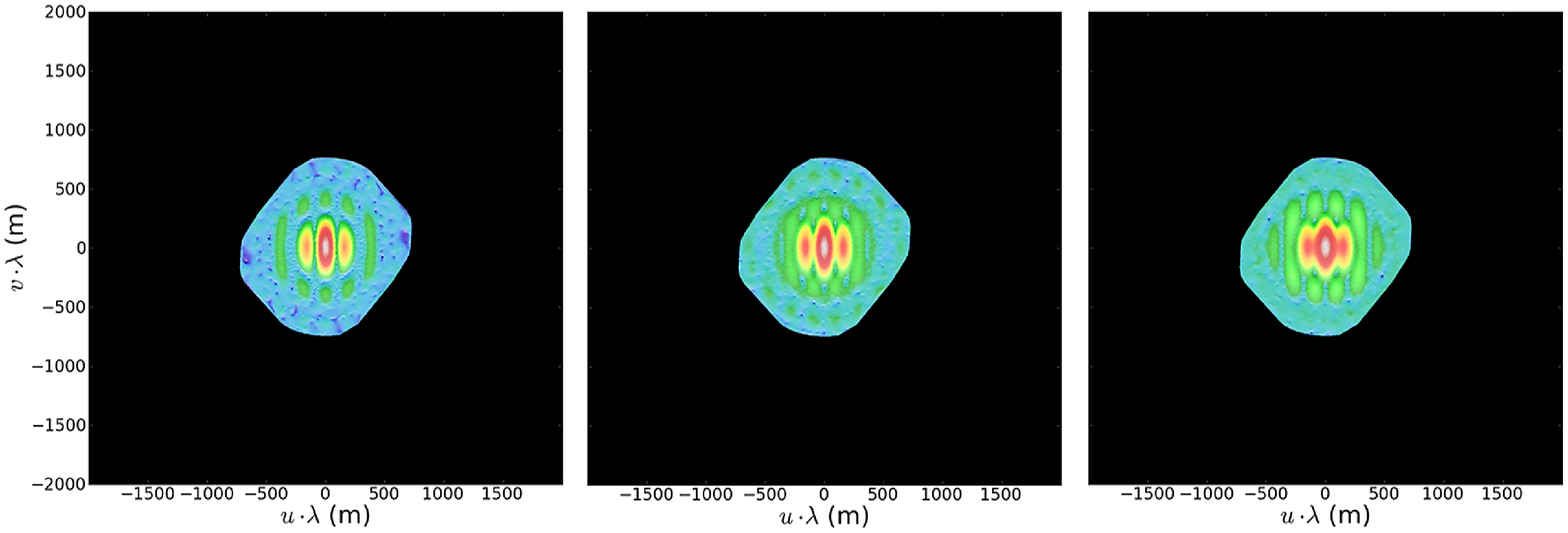}
\caption{Fourier magnitudes in the $(u,v)-$plane, resulting from simulated observations with CTA layout B of binary stars with different diameters.   Already small changes in the diameter of the secondary star by a few tens of {$\mu$}as show up clearly, also well visible in the image reconstructions of Figure \ref{binsep}. }
\label{binsepFT}
\end{figure}

\subsection{Image reconstruction}
\label{images}

Various mathematical methods (of different numerical sophistication and sensitivity to various types of noise) can be applied for the reconstruction of images, and the development of optimum algorithms is a research topic of its own like, perhaps, was the case in early radio interferometry, before today's standard procedures (such as {\it{CLEAN}}) were developed.  Nu{\~n}ez et al.\ \cite{Nunez10,Nunez12a,Nunez12b} applied Cauchy-Riemann based phase recovery to reconstruct images from simulated observations of oblate rotating stars, binary stars, and stars with brighter or darker regions, demonstrating that also rather complex images can be reconstructed on submilliarcsecond scales.  A limitation that remains is the non-uniqueness between the image and its mirrored reflection. 

Figure \ref{binsep} shows the results from such simulations of three binary stars, where the radius of one of them is varied.  Already a change of the diameter of the secondary component by only a few tens of microarcseconds shows up clearly in the Fourier magnitude, and also the reconstructed images reproduce the stellar diameters and separations with quite satisfactory accuracy.

The fidelity of the reconstructed image depends not only on `obvious' factors such as the brightness of the target and the efficiency of the detectors but also on the position of the source in the sky, the geometric layout of the telescope array, and the latitude of the observatory.  Aperture synthesis is achieved by the Earth's rotation carrying the star across the sky and -- since the telescopes are fixed on the ground -- the effective baselines, i.e., the separations between pairs of telescopes as seen along the line of sight, gradually change, filling in various portions of the $(u,v)-$plane.   The geometry of the array and the celestial position of the source determine what projected baselines will be generated during the source's passage across the sky.  For example, sources near the celestial poles do not move, and layout geometries with telescopes in repetitive patterns offer fewer unique baselines.

\subsection{Stellar Diameters and Binary Separations}

The main purpose of the classical interferometer at Narrabri was to measure angular diameters of stars, practical already with only two telescopes.  It was also possible to study parameters such as binary separations by fitting models to the data \cite{HB74}.  With CTA, one will be able to perform such measurements in a much more accurate and model-independent manner, since such an array samples the Fourier plane in thousands of points as compared to the $\sim$5 points for typical past measurements at Narrabri.  For such studies, utilizing some prior knowledge of the source (assuming it to be a binary star, for example), one can fit a model to the `raw' data in the Fourier plane, without going into any (possibly algorithm-dependent) image reconstruction.

\subsection{Observations with subsets of the configurations}

\begin{figure}
\centering
\includegraphics[width=11cm, angle=90]{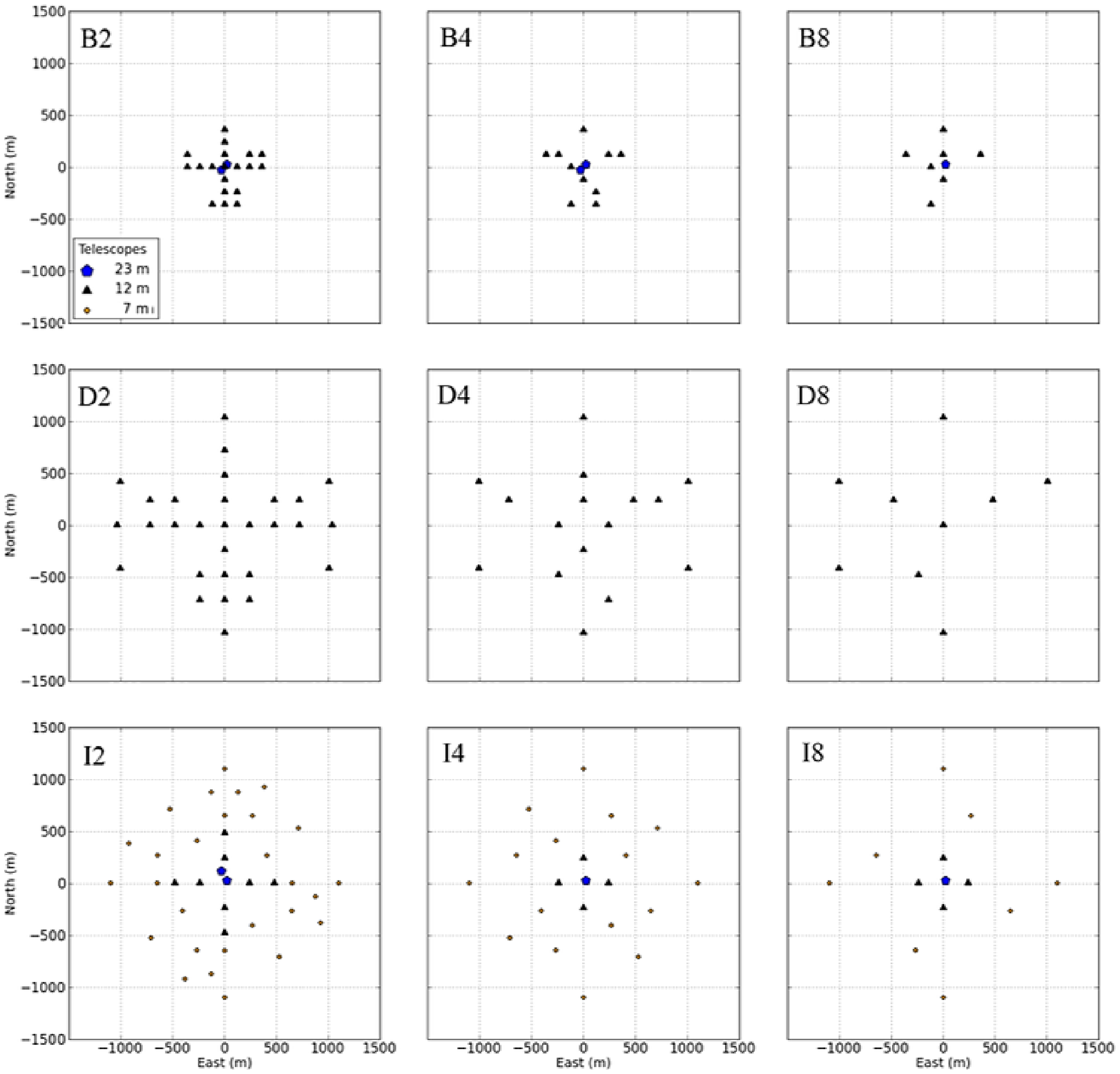}
\caption{Subsets of the candidate configurations (B, D, I from top to bottom). In the leftmost column one half of the telescopes of the superset configurations (Figure \ref{configs}) were selected in a pseudo-random fashion. In the middle column, one in four telescopes was selected, and in the rightmost column one in eight.}
\label{subsets}
\end{figure}

\begin{figure}
\centering
\includegraphics[width=11cm,angle=90]{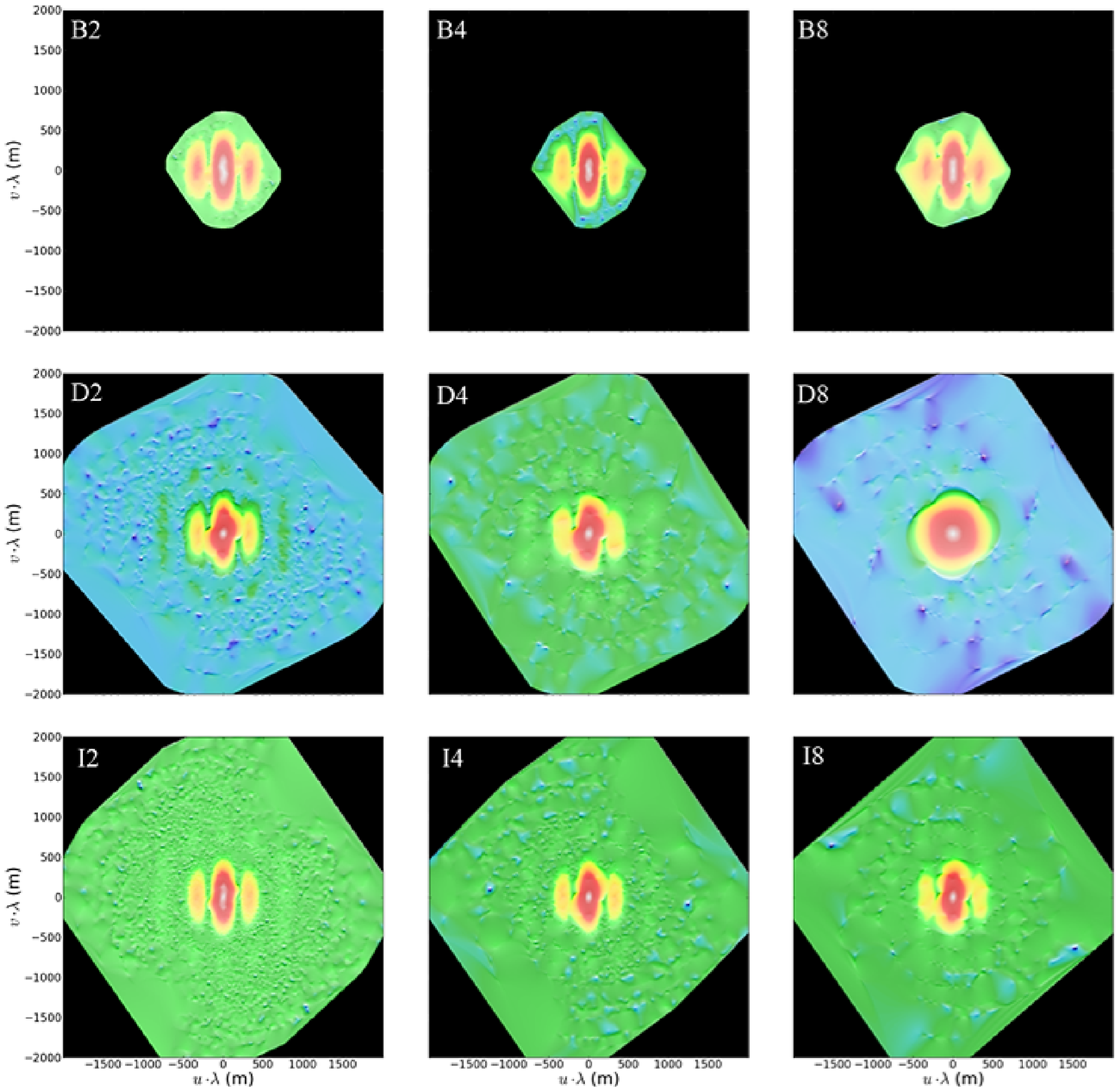} 
\caption{Simulated observations of the binary star in Figure \ref{input} with subsets of configuration B (top row), D (middle row) and I (bottom row).  The left column is for subsets containing half of the telescopes; the center for a quarter, and the rightmost for one eighth of the total (cf.\ Figure \ref{subsets}).  Best imaging is achieved with those arrays that have a balanced mix of different baselines. }
\label{subsetobs}
\end{figure}

Quite probably, not all CTA telescopes will be available for intensity interferometry at any one time.  Even if most of the hardware installed for Cherenkov-light measurements could be utilized for stellar observations, its specific requirements with regard to signal handling (and possibly also auxiliary optics and detectors), combined with finite resources, are likely to limit the number of telescopes initially equipped for interferometry.  Observations using only a subset of the telescopes may thus represent a realistic mode of operation, and we now consider the choice of subsets of telescopes.

For each of the array layout configurations B, D and I (Figure \ref{configs}), three subsets were generated, as shown in Figure \ref{subsets}. The configurations shown in the leftmost column, designated B2, D2 and I2, were obtained by selecting half of the full set of telescopes in a semi-random manner, attempting to preserve the overall `shape' of the array \cite{Jensen10}. In the middle column, one in four telescopes was retained (B4, D4, I4) and in the rightmost column only one telescope in eight was kept (B8, D8, and I8). 

Figure \ref{subsetobs} shows the output from simulations of the binary star in Figure \ref{input} using these subsets. The magnitude of the star was now fixed to $m_V$ =\,5 and a long integration time was chosen in order to depress measurement noise and thus highlight sampling effects for the various configurations.

It is obvious that more numerous telescopes, with a wider distribution of baselines, are better in terms of Fourier-plane sampling, and having only few telescopes restricts the results.  Also, the optimal distribution of baselines depends on the actual size of the target.  In this example, much of the information is at lower spatial frequencies, better sampled by the configuration B8 which for those provides a denser sampling of the $(u,v)-$plane.  Although D8 in principle enables higher angular resolution, its sparse sampling with only a small variation of baseline lengths does not allow the Fourier magnitudes to be correctly estimated at low spatial frequencies, resulting in a somewhat blurred pattern.

Note, however, that here it was not attempted to optimize the telescope selection for optimal sampling.

\section{The new stellar physics} \label{new_stellar}

With optical imaging approaching resolutions of tens of microarcseconds (and with also a certain spectral resolution), we move into novel and previously unexplored parameter domains.  This requires attention not only to optimizing the instrumentation but also to a careful choice of targets to be selected which should be both astronomically interesting and realistic to observe.  With a foreseen brightness limit of perhaps m$_V$=\,6 or 7, and with sources of a sufficiently high brightness temperature, initial observing programs have to focus on bright stars or stellar-like objects \cite{Dravins10}.

Among the about 9000 objects in the Bright Star Catalogue \cite{Hoffleit95}, some 2600 objects are {\it{both}} hotter than 9000\,K {\it{and}} brighter than m$_V$=\,7, among which the brightest and hottest should be those easiest to observe.   A selection of some 35 stars brighter than m$_V$=~2  or hotter than T$_{eff}$~=\,25,000\,K, and of special astrophysical interest were listed as candidates for early observations by Dravins et al.\ \cite{Dravins12}, including the following categories:

\begin{figure}
\centering
\includegraphics[width=7cm, angle=90]{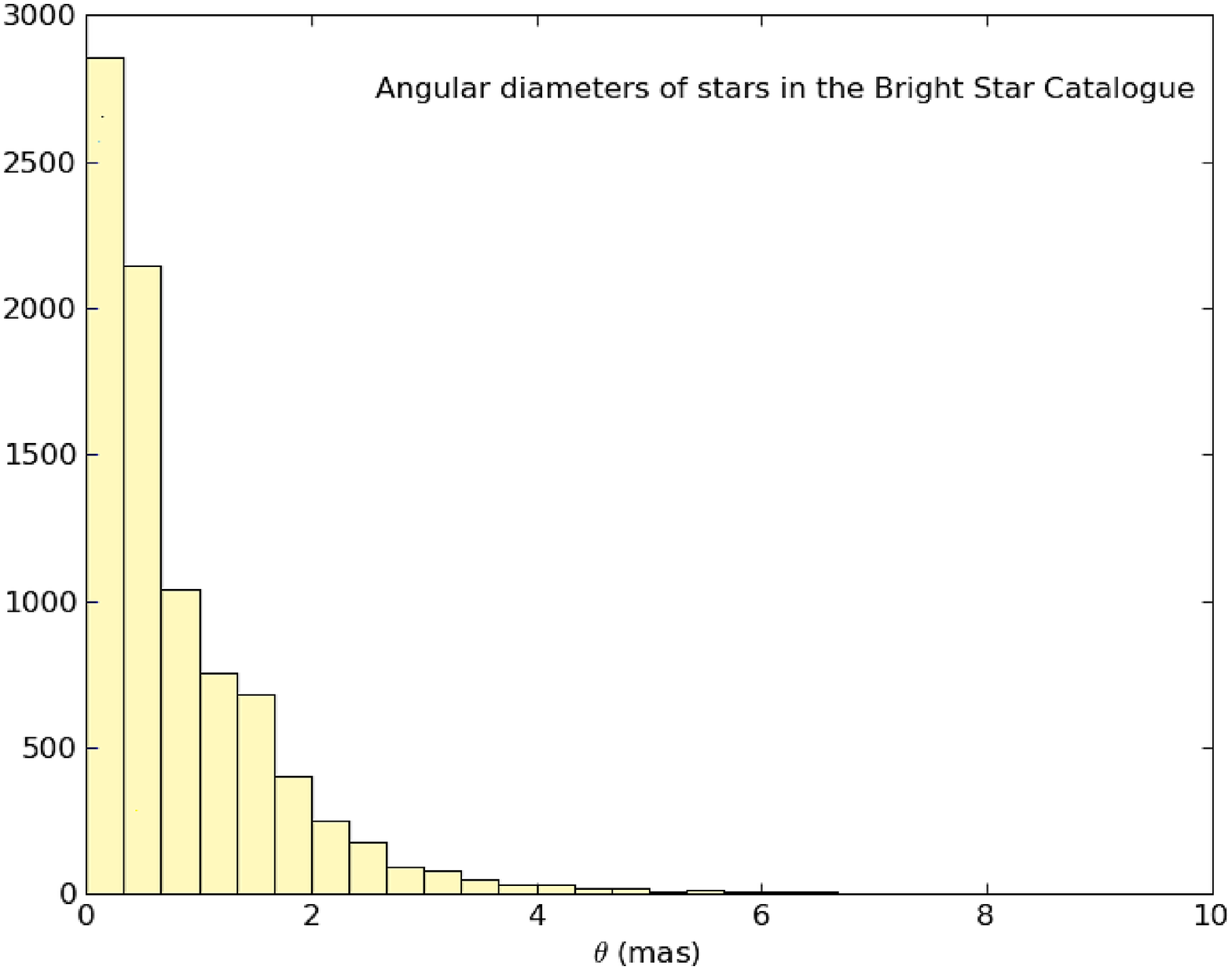}
\caption{Approximated angular diameters of the stars in the Bright Star Catalogue \cite{Hoffleit95}, containing all stars in the sky with m$_V$~$\lesssim$~6.5. Effective temperatures were estimated from the $B-V$ color index listed there, using a polynomial fitted to the relation between that index and T$_{eff}$ by Bessell et al.\ \cite{Bessell98}.  The temperatures were then used to calculate the angular diameters by approximating stars as blackbodies with uniform circular disks.}
\label{diameters}
\end{figure}

\subsection{Rapidly rotating stars} \label{targets_1}

Rapid rotators are normally hot and young stars, of spectral types O, B, and A.  Some are rotating so fast that the effective gravity in their equatorial regions becomes very small, easily enabling mass loss or the formation of circumstellar disks.  Rapid rotation causes the star itself to become oblate, and induces gravity darkening: equatorial regions become dimmer, and polar ones brighter.  

A number of these have now been studied with amplitude interferometers.  By measuring diameters at different position angles, the rotationally flattened shapes of the stellar disks are determined.  For some stars, also their asymmetric brightness distribution across the surface is seen, confirming the expected gravitational darkening and yielding the inclination of the rotational axes.  Aperture synthesis has permitted the reconstruction of images using baselines up to some 300\,m, corresponding to resolutions of 0.5\,mas in the near-infrared H-band around  $\lambda$~1.7\,$\mu$m \cite{Zhaoetal09}.

Two stars illustrate different extremes: Achernar ($\alpha$~Eridani) is a highly deformed Be-star (V$_{rot}$sin i = 250\,km\,s$^{-1}$; $>$~80 \% of critical).  Its disk is the flattest so far observed -- the major/minor axis ratio being 1.56 (2.53 and 1.62\,mas, respectively); and this projected ratio is only a lower value -- the actual one could be even more extreme \cite{Dominic03}.  Further, the rapid rotation of Achernar results in an outer envelope seemingly produced by a stellar wind emanating from the poles \cite{Kervella06,Kervella09}.  There is also a circumstellar disk with H{$\alpha$}-emission, possibly structured around a polar jet \cite{Kanaan08}. The presence of bright emission lines is especially interesting: since the S/N of an intensity interferometer is independent of the spectral passband, studies in the continuum may be combined with observations centered at an emission line.

Going to the other extreme, Vega ($\alpha$~Lyrae, A0 V) has been employed as one of the standard northern hemisphere calibration stars for optical astronomy, but its true spectrum has turned out to be quite complex.  First, space observations revealed an excess flux in the far infrared, an apparent signature of circumstellar dust.  Later, optical amplitude interferometry showed an enormous (18-fold) drop in intensity at $\lambda$~500 nm from stellar disk center to the limb, indicating that Vega is actually a very rapidly rotating star which just happens to be observed nearly pole-on.  The true equatorial rotational velocity is estimated to 270\,km\,s$^{-1}$; while the projected one is only 22 km\,s$^{-1}$ \cite{Aufden06,Peterson06}.  The effective polar temperature is around 10,000\,K, the equatorial only 8,000\,K.  The difference in predicted ultraviolet flux between such a star seen equator-on, and pole-on, amounts to a factor five, obviously not a satisfactory state for a star that should have been a fundamental standard.

\begin{figure}
\centering
\includegraphics[width=10cm, angle=90]{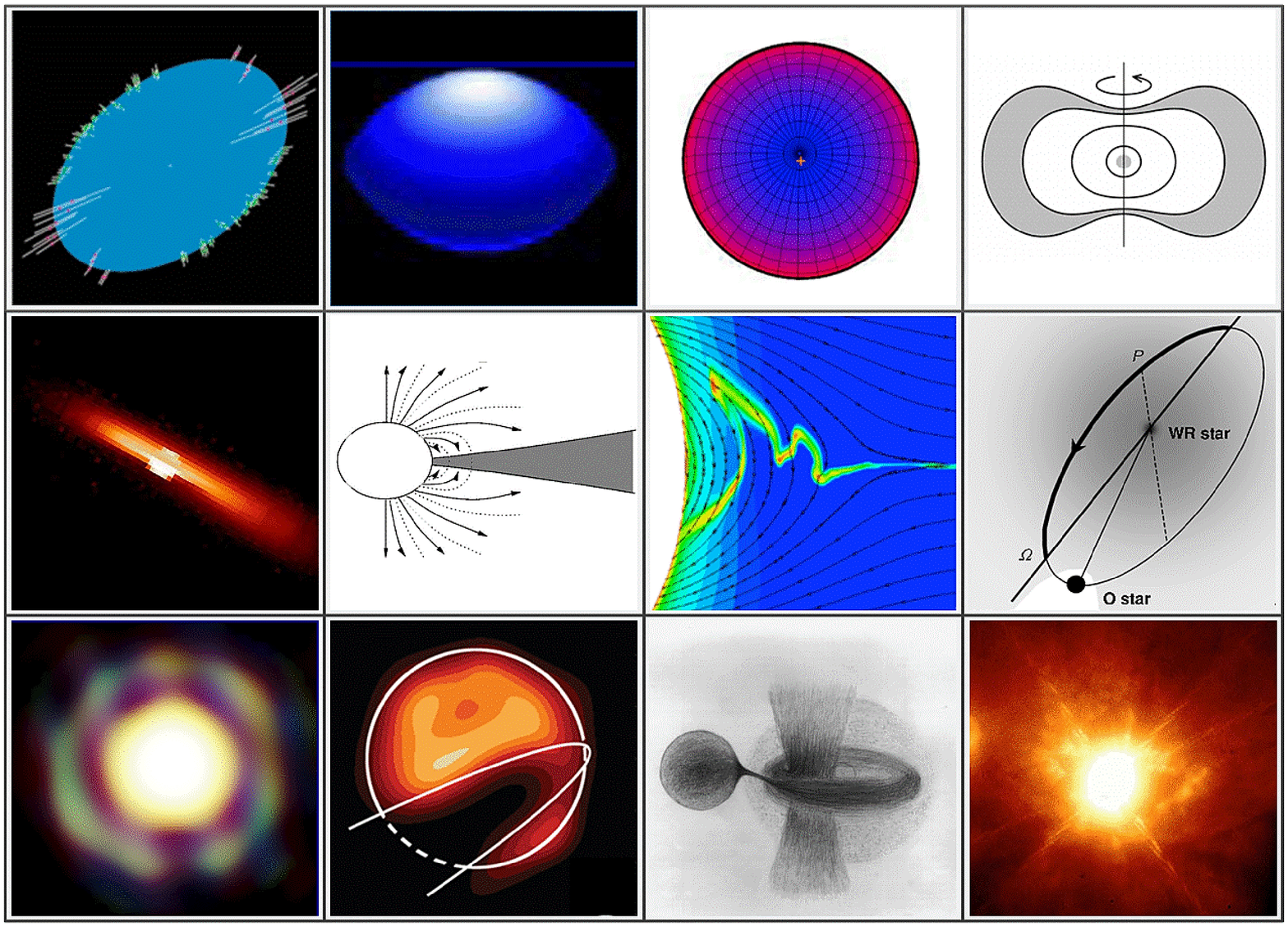}
\caption{Types of primary targets for kilometer-scale intensity interferometry.  Top row: Stellar shapes and surfaces affected by rapid rotation -- The measured shape of Achernar \cite{Dominic03}; expected equatorial bulge and polar brightening of a very rapid rotator \cite{Townsend04}; deduced surface brightness of the rapidly rotating star Vega, seen pole-on \cite{Peterson06}; possible donut-shape for a rapidly and differentially rotating star \cite{MacGregor07}.  Middle row: Disks and winds -- Modeled interferometric image of the circumstellar disk of the Be-star $\zeta$~Tauri \cite{Carciofi09}; a magnetic stellar wind compresses a circumstellar disk \cite{Porter03}; simulation of how stronger magnetic fields distort wind outflow from hot stars \cite{udDoula02}; the strongest stellar wind in a binary opens up cavities around the other star: the geometry around the Wolf-Rayet star $\gamma^2$~Velorum as deduced from interferometry \cite{Millour07}.  Bottom row: Stellar surroundings -- Interferometric image of the giant star T Leporis surrounded by its molecular shell \cite{LeBouquin09}; an analogous image of the giant $\epsilon$~Aurigae, while partially obscured by a circumstellar disk \cite{Kloppenborg10}; artist’s view of the interacting $\beta$~Lyrae system with a gas stream, accretion disk, jet-like structures and scattering halo \cite{Harman02}; an adaptive-optics, high-resolution image of the mysterious object $\eta$~Carinae, the most luminous star known in the Galaxy \cite{ESO03}.}
\label{targets}
\end{figure}

\subsection{Circumstellar disks} \label{targets_2}

Rapid rotation lowers the effective gravity near the stellar equator which enables centrifugally driven mass loss and the development of circumstellar structures.  Be-stars make up a class of rapid rotators with dense equatorial gas disks; the `e' in `Be' denotes the presence of emission in H$\alpha$ and other lines.  Observations indicate the coexistence of a dense equatorial disk with a variable stellar wind at higher latitudes, and the disks may evolve, develop and disappear over timescales of months or years \cite{Porter03}.   The detailed mechanisms for producing such disks are not well understood, although the material in these decretion (mass-losing) disks seems to have been ejected from the star rather than accreted from an external medium.  

\subsection{Winds from hot stars} \label{targets_3}  
     
The hottest and most massive stars (O-, B-, and Wolf-Rayet types) have strong and fast stellar winds that are radiatively driven by the strong photospheric flux being absorbed or scattered in spectral lines formed in the denser wind regions.  Not surprisingly, their complex time variability is not well understood.  Stellar winds can create co-rotating structures in the circumstellar flow in a way quite similar to what is observed in the solar wind. These structures have been suggested as responsible for discrete absorption components observed in ultraviolet P Cygni-type line spectra. 

Rapid stellar rotation causes higher temperatures near the stellar poles, and thus a greater radiative force is available there for locally accelerating the wind.  In such a case, the result may be a poleward deflection of wind streamlines, resulting in enhanced density and mass flux over the poles and a depletion around the equator (opposite to what one would perhaps `naively' expect in a rapidly rotating star).  Surface inhomogeneities such as cooler or hotter starspots cause the local radiation force over those to differ.  This leads to locally faster or cooler stellar-wind streamers which may ultimately collide, forming co-rotating interaction regions.  Further, effects of magnetic fields are likely to enter and -- again analogous to the case of the solar wind -- such may well channel the wind flow in complex ways \cite{udDoula02}.

\subsection{Wolf-Rayet stars and their environments} \label{targets_4}

Being the closest and brightest Wolf-Rayet star, and residing in a binary jointly with a hot O-type star,  $\gamma^2$~Velorum is an outstanding object for studies of circumstellar interactions.  The dense Wolf-Rayet wind collides with the less dense but faster O-star wind, generating shocked collision zones, wind-blown cavities and eclipses of spectral lines emitted from a probably clumpy wind \cite{Millour07, North07}. The bright emission lines enable studies in different passbands, and already with the Narrabri interferometer, Hanbury Brown et al.\ \cite{HB70} could measure how the circumstellar emission region (seen in the C~III-IV feature around $\lambda$~465\,nm) was much more extended than the continuum flux from the stellar photosphere, and seemed to fill much of the Roche lobe between the two components of the binary. 

A few other binary Wolf-Rayet stars with colliding winds are bright enough to be realistic targets.  One is WR 140 (m$_V$\,=\,6.9, with bright emission lines), where the hydrodynamic bow shock has been monitored with milliarcsecond resolution in the radio, using the Very Long Baseline Array (VLBA).  This revealed how the bow-shaped shock front rotates as the orbit progresses during its 7.9\,yr period \cite{Dough05}.

\subsection{Blue supergiants and related stars} \label{targets_5}

Luminous blue variables occupy positions in the Hertzsprung-Russell diagram adjacent to those of Wolf-Rayet stars, and some of these objects are bright enough to be candidate targets, e.g., P~Cyg (m$_V$\,=\,4.8).  Luminous blue variables possess powerful stellar winds and are often believed to be the progenitors of nitrogen-rich WR-stars.  Rigel ($\beta$~Ori; B8~Iab) is the closest blue supergiant (240\,pc). It is a very dynamic object with variable absorption/emission lines and oscillations on many different timescales.  Actually, the properties of Rigel resemble those of the progenitor to supernova SN1987A.

A most remarkable luminous blue variable is $\eta$~Carinae, the most luminous star known in the Galaxy.  It is an extremely unstable and complex object which has undergone giant eruptions with huge mass ejections during past centuries.  The mechanisms behind these eruptions are not understood but, like Rigel, $\eta$~Car may well be on the verge of exploding as a core-collapse supernova.  Interferometric studies reveal asymmetries in the stellar winds with enhanced mass loss along the rotation axis, i.e., from the poles rather than from the equatorial regions, resulting from the enhanced temperature at the poles that develops in rapidly rotating stars \cite{vanBoekel03, Weigelt07}. 

\subsection{Interacting binaries} \label{targets_6}

Numerous stars in close binaries undergo interactions involving mass flow, mass transfer and emission of highly energetic radiation,  Indeed, many of the bright and variable X-ray sources in the sky belong to that category.  However, to be a realistic target for intensity interferometry, they must also be optically bright, which typically means B-star systems.  

One well-studied interacting and eclipsing binary is $\beta$~Lyrae (Sheliak; m$_V$\,=\,3.5).  The system is seen close to edge-on and consists of a B7-type, Roche-lobe filling and mass-losing primary, and an early B-type mass-gaining secondary. This secondary appears to be embedded in a thick accretion disk with a bipolar jet seen in emission lines, causing a light-scattering halo above its poles.  The donor star was initially more massive than the secondary, but has now shrunk to about 3\,M$_\odot$, while the accreting star has reached some 13\,M$_\odot$.  The continuing mass transfer causes the 13-day period to increase by about 20 seconds each year \cite{Harman02}. 

Using the CHARA interferometer with baselines up to 330\,m, the $\beta$~Lyr system has been resolved in the near-infrared H and K bands \cite{Zhaoetal08}.  The images resolve both the donor star and the thick disk surrounding the mass gainer, 0.9\,mas away.  The donor star appears elongated, thus demonstrating the photospheric tidal distortion due to Roche-lobe filling.  Numerous other close binaries invite studies of mutual irradiation, tidal distortion, limb darkening, rotational distortion, gravity darkening, and oscillations.

\subsection{Observing programs} \label{programs} 

Promising targets for early intensity interferometry thus appear to be relatively bright and hot, single or binary O-, B-, and WR-type stars with their various circumstellar emission-line structures.  The expected diameters of their stellar disks are typically on the order of 0.2--0.5\,mas and thus lie (somewhat) beyond what can be resolved with existing amplitude interferometers.  However, several of their outer envelopes or disks extend over a few mas and have already been resolved with existing facilities, thus confirming their existence and providing hints on what types of features to expect when next pushing the resolution by another order of magnitude.  Also, when observing at short wavelengths (and comparing to amplitude interferometer data in the infrared), one will normally observe to a different optical depth in the source, thus beginning to reveal also its three-dimensional structure.

Also some classes of somewhat cooler objects are realistic targets.  Some rapidly rotating A-type stars of temperatures around 10,000\,K should be observable for their photospheric shapes (maybe even watching how the projected shapes change with time, as the star moves in its binary orbit, or if the star precesses around its axis?). 

The exact amounts of observing time required for different targets are somewhat awkward to estimate since -- in contrast to `classical' observations, the achievable signal-to-noise ratio in intensity interferometry depends on several factors other than apparent magnitude (not least the source's own brightness temperature in either the continuum or in some spectral line).  It is also a function of the (normally unknown) source structure: possible high-contrast features on the milliarcsecond level will produce more measurable Fourier power in the $(u,v)-$plane.  Of course, the signal-to-noise improves with higher detective quantum efficiency, better electronic time resolution, the number (and size) of telescopes used, and with the number of wavelength channels that are simultaneously handled; Eq.\eqref{signalnoise}.  

Various simulations (and an extrapolation from past experience with the Narrabri stellar intensity interferometer) have shown \cite{Dravins12,LeBohec06} that a few tens of hours of integration with a large array of the CTA type will faithfully reproduce the Fourier pattern in the $(u,v)-$plane for a star of visual magnitude m$_V$\,=\,5, and T$_{eff}$\,=\,10,000\,K, when measured in one wavelength channel with a time resolution of 1\,ns.  From such data, a two-dimensional image clearly can be reconstructed.  While significant progress on image reconstruction algorithms has recently been made \cite{Nunez12a,Nunez12b}, it is not yet known how sensitive reconstructed images could be to various types of noise levels, either such due to limited source brightness, limited $(u,v)-$plane coverage, or instrumental systematics.  If, in the event of noisy observations, the data would not permit full two-dimensional imaging, one could instead extract one-dimensional quantities such as the sizes of stellar surface structures, the amount of limb darkening, or the separation of binary components.

\section{Observing in practice} \label{practice}

In carrying out actual observations for intensity interferometry, various practical and technical issues may require attention, concerning aspects of the telescopes, detectors, data handling, and the scheduling of observations.

\subsection{Optical $e$-interferometry} \label{e-interfer}

Electronic combination of signals from multiple telescopes is becoming common for long-baseline radio interferometry, where remote radio antennas are connected to a common signal-processing station via optical fiber links in so-called $e$-VLBI.  This is feasible due to the relatively low radio frequencies (MHz-GHz); doing the same for a corresponding optical phase-resolved signal (THz-PHz) would not be possible  but the much slower intensity-fluctuation signals (again MHz-GHz) are realistic to transmit, thus enabling an electronic connection of also optical telescopes.  A number of authors have noted this potential of electronically combining multiple optical apertures, especially for observations at short optical wavelengths (perhaps using the multiple mirror segments of the primary mirror in extremely large telescopes \cite{Dravins08, Dravins05}).  Ofir \& Ribak \cite{Ofir06a, Ofir06b, Ofir06c} evaluate concepts for multidetector intensity interferometers, and even space-based intensity interferometry has been proposed \cite{Hylandetal07, Klein07}, exploiting the possibility to combine signals off-line from each component telescope, thus relaxing the requirement for spacecraft orientation and orbital stability.

\subsection{Performance of Cherenkov telescopes} \label{ACT_issues}

The signals to be measured for intensity interferometry have much in common to those of atmospheric Cherenkov flashes: nanosecond time structure and relatively short optical wavelengths.  Most probably, the same types of very fast photon-counting detectors can be used, although the sources to be observed are much brighter, and the data handling has to allow for continuous integrations (rather than trigger-based acquisition of short data bursts). 

\subsubsection{Image quality} \label{image_quality}

Even if the technique of intensity interferometry as such does not require good optical quality, and permits also rather coarse light collectors with point-spread functions of several arcminutes, issues arise from unsharp stellar images: in particular an increased contamination from the background light of the night sky.  Although this light does not contribute any net intensity-correlation signal, it increases the photon-counting noise, especially when observing under moonlight conditions.

While any reasonable optical quality should be adequate for intensity interferometry as such, the magnitude m$_V$ of the faintest stars that can be studied will be influenced by the optical point spread function.  Two extreme sky brightness situations can be: (a) dark observatory sky with $\sim$\,21.5\,m$_V$/arcsec$^2$; (b) sky with full Moon; $\sim$\,18\,m$_V$/arcsec$^2$.  The equivalent magnitudes from the sky background then result in m$_V$\,$\sim$\,9.4 (a) and 5.9 (b) for a 5\,arcmin diameter field, and m$_V$\,$\sim$\,12.9 (a) and 9.4 (b) for a 1\,arcmin diameter field.

A larger point spread function also takes in other sky events (meteors, distant flashes of lightning, etc.), and may preclude the use of small-sized semiconductor detectors of possibly higher quantum efficiency.

\subsubsection{Isochronous optics} \label{isochronous}

For Cherenkov light observations, a large field of view is desired.  In most optical systems, the image quality deteriorates away from the optical axis, and to mitigate this, various optical solutions are used.  Many current telescopes have the layout introduced by Davies \& Cotton \cite{Davies57}, whose primary reflector forms a spherical structure centered on the focal point, giving smaller aberrations off the optical axis compared to a parabolic design. 

This has the consequence that the telescope optics become anisochronous, i.e., photons originally on the same wavefront, but striking different parts of the entrance aperture may not arrive to the focus at exactly the same time.  As noted above, the signal-to-noise ratio improves with electronic bandwidth, i.e., the time resolution with which stellar intensity fluctuations can be measured.  The time spread induced by anisochronous telescopes acts like `instrumental broadening' in the time domain, filtering away the most rapid fluctuations.  This probably is not a serious issue since the gamma-ray induced Cherenkov light flashes in air last only a few nanoseconds, and thus the performance of Cherenkov telescopes cannot be made much worse, lest they would lose sensitivity to their primary task.  Still, since realistic electronics may reach resolutions on the order of 1 ns, it would be desirable that the error budget does not have components in excess of such a value.

Among existing Cherenkov telescopes, this is satisfied by parabolic designs (e.g., MAGIC) but not by the Davies-Cotton concept (e.g., VERITAS or H.E.S.S.-I).  For example, in the H.E.S.S.-I telescopes the photons are spread over $\Delta$t\,$\sim$\,5 ns, with an rms width $\sim$\,1.4\,ns \cite{Ahper04,Bernlohr08,Schliesser05}.  For large telescopes, the time spread would become unacceptably large if a Davies-Cotton design were chosen, and those therefore normally are parabolic (e.g., MAGIC on La Palma; H.E.S.S.-II in Namibia, and MACE in Ladakh, India).  In principle, these then become isochronous -- apart from minute (few hundred ps) effects caused by individual mirror facets being spherical rather than parabolic, or by the tesselated mirror facets being mounted somewhat staggered in depth. 

Also non-parabolic telescopes can be made effectively isochronous, if they have more than one optical element.  The two-mirror Schwarzschild-Couder design is attractive for smaller telescopes, not least because its smaller image scale permits smaller and less expensive focal-plane cameras \cite{Vassilev07}.  For on-axis rays, this design in principle is isochronous, but the time spread increases to $\sim$\,1\,ns for angles a few degrees off center.  Also, Schmidt-type telescope designs may satisfy high demands on isochronicity (even better than 10\,ps on axis), while also being compact, offering a wide field of view, and having a narrow point-spread function \cite{Mirzoyan09}.  However, to take full advantage of such performance would require corresponding accuracies in all other components of the error budget, including the signal handling, and the positioning of telescopes on millimeter scales.  Such values also begin to approach the level of natural fluctuations in path-length differences induced by atmospheric turbulence \cite{Cavazzani12,Wijaya11}.

\subsubsection{Focusing at `infinity'} \label{focus}

The optical foci of Cherenkov telescopes are optimized to correspond to those heights in the atmosphere where most of the Cherenkov light originates, and the image of a distant star will then be slightly out of focus.  For a focal length of $f$\,=\,10\,m, the focus shifts 1\,cm between imaging at 10\,km distance and at infinity, which for an $f$/1 telescope implies an additonal image spread of some cm.  In order to decrease the stellar image and not to take in too much of the night-sky background, it is desirable (though not really mandatory) to refocus the telescope on stars at `infinity'.   On some (especially larger) telescopes, such a possibility may be available anyhow since some refocusing can be required in response to mechanical deflections when pointing in different elevations or as caused by nocturnal or seasonal temperature variations.  In the absence of such a possibility, a refocusing could still be achieved by an optical lens placed directly in front of the photosensor.

\subsubsection{Placement of telescopes in an array} \label{layout}

The placement of telescopes in interferometers can be optimized for the best coverage of the $(u,v)-$plane \cite{Boone01,Herrero71,Holdaway99, Keto97, Mugnier96, Thompson01}.  As the star gradually crosses the sky during the night, projected baselines between pairs of telescopes change, depending on the angle under which the star is observed.  If the telescopes are placed in a regular geometric pattern, e.g., a repetitive square grid, the projected baselines are similar for many pairs of telescopes, and only a limited region of the $(u,v)-$plane is covered (on the other hand, redundant baselines result in better signal-to-noise for those particular ones).  Since stars rise in the east, moving towards west, baselines between pairs of telescopes that are not oriented exactly east-west will trace out a wider variety of patterns.  Because of such considerations, existing amplitude interferometers (both optical and radio) locate their component telescopes in some optimal manner (e.g., in a Y-shape, or in logarithmic spirals, unless constrained by local geography). 

As concerns specifically the CTA, its smaller telescopes will be so numerous that, for most practical purposes, their exact placement should not be critical for interferometry -- a huge number of different baselines will be available anyway.  However, the situation is different for the very few large telescopes.  Avoiding placing them on a regular grid (such as a square) will offer a variety of baseline lengths, give a better coverage of the $(u,v)-$plane, and permit better image reconstruction.

\subsubsection{Impact on observatory operations} \label{operations}

The impact of intensity interferometry on other Cherenkov array operations should not be significant.  An important aspect is that -- while full moonlight may constrain observations of the feeble atmospheric Cherenkov light -- measuring brighter stars is no problem for intensity interferometry, enabling efficient operations during both bright- and dark-Moon periods.

Potential sources for interferometry are distributed over large parts of the sky and permit vigorous observing programs from both northern and southern sites.  However, several among the hot and young stars belong to Gould's Belt, an approximately 30 million year old structure in the local Galaxy, sweeping across the constellations of Orion, Canis Major, Carina, Crux, Centaurus, and Scorpius, centered around right ascensions 5-7 hours, not far from the equator.  Thus, many primary targets are suitable to observe during northern-hemisphere winter or southern-hemisphere summer.  We note that this part of the sky is far away from the many gamma-ray sources near the center of the Galaxy (which is at right ascension 18 hours).

\subsection{Detectors and cameras} \label{detector_issues}

Typical Cherenkov telescopes have focal lengths on the order of 10\,m, providing a focal-plane image scale around 3\,mm/arcmin.  A typical point-spread function of 3\,arcmin diameter thus corresponds to 1\,cm.  Detectors that are capable of photon counting with nanosecond time resolution include well-established vacuum-tube photomultipliers and large-size solid-state avalanche diode arrays that are under development.

A Cherenkov telescope typically holds several hundred photomultiplier tubes acting as `pixels' in its focal-plane camera.  The detectors and their ensuing electronics are naturally optimized for the triggering on, and the recording of, faint and brief transients of Cherenkov light and might not be readily adaptable for hour-long continuous recordings of bright stellar sources.  However, for intensity interferometry, only one pixel is required (at least in principle, although some provision for measuring the signal at zero baseline is required) and we note that in some telescopes (e.g., HEGRA \cite{OnaWilhelmi04} and MAGIC \cite{Lucarelli08}), the central camera pixel was specifically designed to be accessible for experiments without affecting any others.  Such types of central pixels could be usable to perform some experiments towards also intensity interferometry.

However, even if a special pixel is accessible, it may not be possible to use it in its bare form.  If observing a bright source in broadband white light with a large telescope, the photon-count rate may become too large to handle, even for reduced photomultiplier voltages.  As discussed above, the signal-to-noise ratio in intensity interferometry is independent of the optical passband: the smaller photon flux in a narrow spectral segment is compensated by the increased temporal coherence of the more monochromatic light.  This property can be exploited with some color filter to reduce the photon flux to a suitable level, or using a narrow-band filter tuned to some specific spectral feature of astrophysical significance.  For such uses, there should be some provision for a mechanical mounting in front of the detector to hold some small optical element(s).  A broader-band color filter could simply be placed immediately in front of a photomultiplier but a narrow-band filter could require additional arrangements.  Most such filters are interferometric ones and need to be used in collimated (parallel) light in order to provide a more precise narrow passband.  Since light reaching the Cherenkov camera is not collimated, some additional optics could be required, or else one might use narrow-band filters based on other optical principles, such as Christiansen filters \cite{Bala92}.  

The further development and optimization of observational techniques is likely to involve experiments with other types of detectors, color filters, polarizers or other optical components which could be awkward to mechanically and electronically (re)place in the regular Cherenkov camera.  To minimize disturbances to the Cherenkov camera proper, it could then be preferable to place an independent detector unit on the outside of its camera shutter lid.  Such constructions have already been made on existing Cherenkov telescopes, e.g., a 7-pixel unit on a H.E.S.S. telescope used a plane secondary mirror to put it into focus, and was used for experiments in very high time-resolution optical observations.  Its central pixel recorded the light curve of the target, while a ring of six surrounding pixels monitored the sky background and acted as a veto system to reject atmospheric background events \cite{Deil08, Deil09, Hinton06}.  For such devices, provision must also be made for electrical power supply and signal cables to/from the outside of these camera shutter lids.

\subsection{Signal handling} \label{signal_issues}

Current electronic units, used in various photon-counting experiments, have time resolutions approaching 1\,ns, and the error budget should ideally not have components in excess of such a value (the signal-to-noise is proportional to the square root of the signal bandwidth).  Telescopes may be separated by up to a kilometer or two, and the timing precision of the photon-pulse train from the detector to a central computing location should be assured to no worse than some nanosecond (for the timing of its leading pulse-edge; the pulse-width may be wider).  Such performance appears to be achievable by analog signal transmission in optical fibers \cite{Rose00, White08}.  Compared to metal cables, these have additional advantages of immunity to cross-talk and to electromagnetic interference, and also avoid the difficulty of maintaining a common ground and protection for the receiving electronics against (in some locations not uncommon) lightning strikes across the array.  

Another possibility is using clocks on satellite positioning systems such as GPS or {\it{Galileo}}, where absolute timing within some nanosecond has been achieved in astronomical instrumentation \cite{Naletto09}.  This enables to time-tag the photon stream for later off-line analysis, with an accuracy better than the anisochronicity of the mirror surfaces of the typical Davies-Cotton optical design.

\subsubsection{Correlators} \label{correlators}

A critical element of an intensity interferometer is the correlator which provides the averaged product of the intensity fluctuations $\langle\Delta I_1 \Delta I_2\rangle$ to be normalized by the average intensities $\langle I_1\rangle$ and $\langle I_2\rangle$ (Eq.\ \ref{intcorr4}).  The original interferometer at Narrabri used an analog correlator to multiply the photocurrents from its phototubes, and significant efforts were made to shield the signal cables from outside disturbances.  Current techniques, such as FPGA ({\it{Field Programmable Gate Arrays}}), permit to program electronic units into high-speed digital correlators with time resolutions of a few ns or better.  

Similar units are also commercially available for primary applications in light scattering against laboratory specimens \cite{Saleh78}.  Such intensity-correlation spectroscopy is the temporal analog to the [spatial] intensity interferometry, and was developed after its subsequent theoretical understanding.  It was realized that high-speed photon correlation measurements were required and electronics initially developed in military laboratories were eventually commercialized, first by Malvern Instruments in the U.K. \cite{Pike79}, and nowadays offered by various commercial companies \cite{Becker05}.

An alternative approach (at least for limited photon-count rates) is to digitize and store all data, and then later perform the correlation analyses off-line.  The data streams from multiple telescopes can then be cross-correlated using a software correlation algorithm, permitting the application of digital filtering to eliminate possible interference noise from known sources, and also to compute other spatio-temporal parameters, such as higher-order correlations between three telescopes or more, which in principle may contain additional information.  On the other hand, this requires a massive computing effort and possible observational problems may not get detected while observations are in progress but only at some later time.  Such a capability was foreseen in the design study for {\it{QuantEYE}}, a proposed very high-time resolution instrument for extremely large telescopes \cite{Dravins05, Dravins06}, and verified in the construction and operation of the {\it{AquEYE}} and {\it{IquEYE}} instruments, the latter used also at the European Southern Observatory in Chile \cite{Naletto09, Naletto07, Naletto10}.

\subsection{Delay units} \label{delays} 

Besides the correlator, another piece of electronics is required for real-time intensity interferometry, namely to implement a continuously variable time delay that compensates for the relative timing of the wavefront at the different telescopes, as the source moves across the sky (Eq.\ \ref{rotationsynth}). 

If such a delay unit is not used, the maximum correlation signal in a multichannel digital correlator will appear not in the channel for zero time delay between any pair of telescopes, but rather at that channel which corresponds to a delay equal to the light-time difference between telescopes along the line of sight towards the source.  This arrangement is feasible already with existing digital correlators since these can be programmed to measure the correlation at full time resolution at time coordinates away from zero.

Such arrangements, however, are not required in the case of off-line data analysis, where the delays can be introduced by software afterwards.

\section{Experimental work} \label{experiments}

As preparatory steps towards realizing full-scale stellar intensity interferometry, different laboratory and field experiments have been carried out at various institutes, in particular at the StarBase facility in Utah \cite{Dravins12, LeBohec10, StarBase12}.  Also, in a first full-scale test with a Cherenkov telescope array, pairs of the 12~m telescopes of the VERITAS array in Arizona were used to observe a number of stars, with pairs of its telescopes interconnected through digital correlators \cite{DravinsLeBohec08}.  For these observations, starlight was detected by a photon-counting photomultiplier in the central pixel of the regular Cherenkov-light camera, the outgoing photon pulses were digitized using a discriminator, then pulse-shaped and transmitted from each telescope via an optical cable to the control building where they entered a real-time digital cross correlator, computing the cross correlation function for various time delays.  Continuous count rates up to some 30 MHz were handled, limited by the digitization and signal-shaping electronics.  While these experiments were not intended to measure astrophysical quantities but to gain experience in operating with a full-scale observatory, they confirmed that no fundamental problems seem to exist in carrying out such operations.

\section{Further possibilities} \label{possibilities}

The availability of very large light-collecting areas, distributed over an extended array enables further classes of optical observations, not feasible with ordinary instrumentation.

\subsection{Higher-order spatio-temporal correlations} \label{higher-order}

The quantum theory of optical coherence \cite{Glauber07, Mandel95} describes how one can define correlations between arbitrarily many spatial and/or temporal coordinates in the volume of light (`photon gas') being received from a source.  The spatial intensity interferometer is only one special case of such more general spatio-temporal correlations, in that it measures the cross correlation between the intensities at {\it{two}} spatial locations, at {\it{one}} instant in time.   

However, using telescope arrays, and given that their photon detectors provide data streams which can be analyzed at will, one can construct, e.g., third-order intensity correlations, $g^{(3)}$, for systems of three telescopes: $\langle I(r_1, t_1) I(r_2, t_2) I(r_3, t_3) \rangle$, where the temporal coordinates do not necessarily have to be equal.  In principle, such and other higher-order spatio-temporal correlations in light may carry additional information about the source from where the light has been emitted and thus -- at least in principle -- is of relevance for astronomy where information about the source has to be extracted from more or less subtle properties of its radiation received \cite{Jain08, Ofir06a}.

Although, in the recording of higher-order correlations, also the relative noise level increases (possibly demanding very large telescopes for certain measurements \cite{Dravins94}), all sorts of higher-order correlations can in principle be obtained without any additional observational effort if the digital signals from each telescope are available for further manipulation in either hard- or software.  For example, one could calculate correlations among also all possible triplets and quadruplets of telescopes, possibly enabling a more robust full reconstruction of the source image \cite{Ebstein91, Fontana83, Hyland05, Marathay94, Sato78, Sato81, Schulz98, Zhilyaev08}.

\subsection{High-time-resolution astrophysics}

Further uses of CTA can be envisioned in searching for extremely rapid (micro- or even nanosecond) optical variability, such as suspected from pulsars or other compact sources.  In a few radio pulsars, nanosecond pulse structure has been observed, representing the currently most rapid fluctuations seen in astronomical sources but so far lacking any credible explanation (perhaps nonlinear plasma turbulence, stimulated Compton scattering, or angular beaming due to relativistic source motion?), and there is some evidence that corresponding events may exist also in the optical \cite{Shearer03}.  Despite the point-spread function extending over arcminutes (and thus giving a considerable contribution from the night-sky background), the very large collecting area distributed over several independent telescopes would make such searches more sensitive than with any existing large telescope.  Although the number of photons per second in the background skylight may be very significant, their number per microsecond is still very modest and the sensitivity to detecting the very shortest fluctuations becomes mainly a function of light-collecting area \cite{Lacki11}. 

The sensitivity of Cherenkov telescopes to detecting such very brief optical flashes was analyzed by Deil et al.\ \cite{Deil09}.  Comparing a H.E.S.S. 100\,m${^2}$ Cherenkov telescope with a sky-background-free optical telescope for very high-time-resolution photometry, they found that for flashes shorter than some 100\,ns, a large Cherenkov telescope outperforms today's largest astronomical telescopes, at least under dark-Moon sky conditions.  Optical observations of the Crab pulsar have been made with Cherenkov telescopes of H.E.S.S. \cite{Hinton06} and MAGIC \cite{Lucarelli08}.  Although this optical pulse is only some 10$^{-6}$ of the background (the sky plus the Crab nebula surrounding the pulsar), and integrations over many hundreds of pulsar periods are required before a sensible signal appears, statistical information on also the very fine time structure can be retrieved within very reasonable integration times.

\section{Outlook} \label{outlook}

Interferometry for the attempted measurement of stellar diameters appears to have been first carried out in the 1870's by {\'E}douard St{\'e}phan \cite{Stephan1874}, following a suggestion by Hippolyte Fizeau.  A two-aperture mask was placed over a 80 cm reflector at Marseille Observatory, but it was soon realized that stars could not be resolved over this short baseline.  In the 1920's, Albert Michelson \& Francis Pease \cite{Michelson21} operated a 6-meter interferometer mounted on the 100-inch Hooker telescope on Mt.Wilson, and succeeded in measuring diameters of a few giant stars, while their later 15-meter instrument proved mechanically too unstable for practical use \cite{Hariha85}. 

The demanding requirement to maintain stable optical path differences during observations within a fraction of an optical wavelength caused the technique to lay dormant for half a century, until Antoine Labeyrie \cite{Labeyrie75} succeeded in measuring interference fringes between two separated telescopes.  This success triggered the construction of a whole generation of optical amplitude interferometers and is also said to have been the specific reason why the plans to build a successor to the original Narrabri intensity interferometer (designed around that very time) were not realized, and (as far as astronomy is concerned), the technique has now been dormant for decades.

However, the progress in instrumentation and computing technology since the days of the Narrabri interferometer has been extraordinary. High-speed photon-counting detectors and hardware correlators are commercially available, and new mathematical algorithms allow for image reconstruction. The most valuable components -- large light collectors -- are being realized in the form of air Cherenkov telescopes.  All of this has sparked a renewed interest in astronomical intensity interferometry, and a first workshop (since very many years) on this topic was held not long ago \cite{LeBohec09}.  Thus, long after the pioneering experiments by Hanbury Brown and Twiss, the technological developments carry the promise of achieving a basic but difficult goal: to finally be able to view our neighboring stars not only as mere unresolved points of light but as the extended and most probably very fascinating objects that they really are.

Acknowledgments

The work at Lund Observatory is supported by the Swedish Research Council and The Royal Physiographic Society in Lund.  Stephan LeBohec acknowledges support from grants SGER \#0808636 of the National Science Foundation.  Within the CTA design study, valuable comments about its possible use for intensity interferometry were received from several colleagues, including Michael Andersen, Cesare Barbieri, Isobel Bond, Stella Bradbury, Hugues Castarede, Michael Daniel, the late Okkie de Jager, Willem-Jan de Wit, Wilfried Domainko, C{\'{e}}dric Foellmi, Werner Hofmann, Richard Holmes, David Kieda, Antoine Labeyrie, Lennart Lindegren, Hans-G{\"{u}}nter Ludwig, Giampiero Naletto, Aviv Ofir, Guy Perrin, Andreas Quirrenbach, Erez Ribak, Stephen Ridgway, Joachim Rose, Diego F.\ Torres, Gerard van Belle, and Luca Zampieri.  We also acknowledge critical reviews of the manuscript by SAPO, the CTA Speakers' and Publication Office, and by an anonymous referee.  The research leading to these results has received funding from the European Union's Seventh Framework Programme ([FP7/2007-2013] [FP7/2007-2011]) under grant agreement No.\ 262053.







\end{document}